\documentclass[prl,aps,superscriptaddress,footinbib,twocolumn]{revtex4-2}
\usepackage{color}
\usepackage{amsmath}
\usepackage{amssymb}
\usepackage{graphicx}
\usepackage{siunitx}
\usepackage[unicode=true,
 bookmarks=true,bookmarksnumbered=false,bookmarksopen=false,
 breaklinks=false,pdfborder={0 0 1},backref=false,colorlinks=true,
 hyperfootnotes=false]
 {hyperref}
\hypersetup{
 linkcolor=magenta, urlcolor=blue, citecolor=blue, pdfstartview={FitH}}

\usepackage{amsfonts}
\usepackage{bm}
\usepackage{times}
\hypersetup{urlcolor=blue}

\def\ket#1{\left|#1\right\rangle}
\def\bra#1{\left\langle#1\right|}
\def\braket#1{\left\langle#1\right\rangle}
\newcommand{\bx}{\bm{x}}

\begin{document}


\title{Matter-Mediated Entanglement by Classical Gravity under Realistic Matter Dynamics}

\author{Ziqian Tang}\thanks{These authors contributed equally to this work}
\affiliation{Beijing Key Laboratory of Fault-Tolerant Quantum Computing, Beijing Academy of Quantum Information Sciences, Beijing 100193, China}
\affiliation{Beijing National Laboratory for Condensed Matter Physics, Institute of Physics, Chinese Academy of Sciences, Beijing 100190, China}
\affiliation{University of Chinese Academy of Sciences, Beijing 100049, China}

\author{Chen Yang}\thanks{These authors contributed equally to this work}
\affiliation{School of Physics, Peking University, Beijing 100871, China}

\author{Hanyu Xue}\thanks{These authors contributed equally to this work}
\affiliation{Department of Physics, Massachusetts Institute of Technology, Cambridge, MA 02139, USA}

\author{Haochen Yu}
\affiliation{School of Physics, Peking University, Beijing 100871, China}

\author{Zizhao Han}
\affiliation{Center for Quantum Information, IIIS, Tsinghua University, Beijing 100084, China}

\author{Zikuan Kan}
\affiliation{School of Physics, Renmin University of China, Beijing 100872, China}

\author{Yulong Liu}
\email{liuyl@baqis.ac.cn}
\affiliation{Beijing Key Laboratory of Fault-Tolerant Quantum Computing, Beijing Academy of Quantum Information Sciences, Beijing 100193, China}

\date{\today}

\begin{abstract}
Gravity-induced entanglement (GIE) is widely regarded as key evidence of nonclassical gravity. Recent work, however, argues that a classical gravitational background can generate entanglement through virtual matter propagation between separated masses. Here we show that this mechanism is negligible once realistic matter dynamics is taken into account. For bound or condensed matter, the binding potential converts the claimed long-range matter-mediated interaction into an exponentially short-ranged one, with an attenuation length of only a few picometers for representative binding energies—well below atomic dimensions and therefore negligible under experimentally relevant distances. For unbound matter, the channel remains negligible before significant wave-packet overlap, while after overlap the packets no longer represent two spatially separated subsystems. Thus the proposed classical matter-mediated mechanism does not challenge the usual interpretation of GIE as evidence for nonclassical gravity once physical localization and finite-time matter dynamics are treated consistently.

\end{abstract}

\maketitle



The question of whether gravity can generate entanglement has become a central theme in the search for quantum signatures of the gravitational field. In standard proposals, gravitationally induced entanglement between massive systems is widely regarded as a witness of nonclassical gravity, since a purely classical mediator acting via local operations and classical communication (LOCC) cannot create such correlations \cite{Feynman1957, PhysRevLett.119.240401, PhysRevLett.119.240402, krisnanda2020observable, Biswas2023, bose2025massive, marletto2025quantum}. This paradigm has recently been challenged by the claim that even a classical gravitational background can induce entanglement when matter is treated as a quantum field, through virtual propagation of matter particles between spatially separated localized modes of masses \cite{aziz2025classical}. If appreciable in realistic systems, such a channel would constitute a loophole in the usual interpretation of GIE experiments.

Several subsequent analyses have challenged the claim in \cite{aziz2025classical} from complementary perspectives. Some have questioned the robustness of the claimed entangling contribution, showing that its presence can depend sensitively on how the full dynamics and perturbative expansion are treated, including the factorization of the evolution and the inclusion of same-order processes \cite{diosi2025no,marletto2025classical,gundhi2026classical}. Other arguments emphasize that, if a crossed contribution survives, the resulting entangling effect should be attributed to the matter sector rather than to classical gravitational degrees of freedom \cite{marletto2025local,schneider2025demonstration}. More specifically, within the perturbative QFT framework of the original proposal, this contribution has been identified with matter propagation between the two systems \cite{vidal2026matter}. While these studies address important questions about the existence and attribution of the claimed entanglement, a deliberately conservative question is whether the matter-propagation channel, even if retained, can appreciably affect the interpretation of GIE experiments under the physical conditions required to realize them.

In this work, we show that this apparent loophole in \cite{aziz2025classical} disappears once realistic matter dynamics is taken into account. We identify the relevant entangling contribution with the off-diagonal element of the dressed matter propagator between spatially separated modes, providing a unified framework for bound and unbound matter. For bound or condensed matter, the necessary inclusion of binding potentials converts the claimed long-range matter-mediated interaction into an exponentially short-ranged one, with an attenuation length of only a few picometers for representative binding energies. For unbound finite-energy wave packets, the same channel reduces to ordinary finite-time matter propagation. It remains negligible while the wave packets represent spatially separated subsystems, whereas appreciable mode-to-mode propagation requires substantial overlap, at which point this subsystem interpretation no longer applies.

Furthermore, we show that the previously claimed long-range power-law contribution~\cite{aziz2025classical} arises from combining two incompatible assumptions: taking the infinite-time limit of matter propagation while continuing to treat the wave packets as spatially separated localized modes. In this limit, unbound wave packets inevitably spread and overlap, so the resulting stationary power-law behavior no longer describes an entangling channel between two spatially separated physical subsystems.

Our results therefore rule out the proposed virtual-matter mechanism as an appreciable long-range entangling channel between localized systems in a prescribed classical gravitational background. Once physical localization and finite-time matter dynamics are treated consistently, this apparent loophole disappears, leaving the interpretation of gravitationally induced entanglement as evidence for nonclassical effects of gravity unaffected by this mechanism.

\emph{Mode-propagator formulation.---}
We begin by reformulating the proposed ``virtual-matter exchange'' mechanism to isolate its essential field-theoretic content. Rather than interpreting individual diagrams, we express the effect strictly in terms of matter propagation amplitudes dressed by a classical gravitational background.

\begin{figure}[t]
    \centering
    \includegraphics[width=0.98\linewidth]{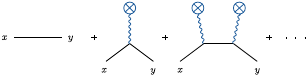}
    \caption{Dyson Series of the dressed matter two-point function $G_g (x,y)$ in the presence of external Newtonian background. Solid lines and curly lines denote matter propagators and gravity insertions respectively, with the crossed circles representing outer current injection. The outer current originated from classical gravity fields in the interaction picture, while restriction potential for bounded matter can also be taken into account in Furry picture.}
    \label{fig:placeholder}
\end{figure}

Consider a quantum matter field $\hat{\phi}(x)$ evolving in the presence of a prescribed classical spacetime perturbation $h_{\mu\nu}(x)$, which is treated as a $c$-number field. The interaction is governed by the Hamiltonian $\hat{H}^{\rm CG} = (-1/2)\int d^3\mathbf{x}\, h_{\mu\nu}(x) \hat{T}^{\mu\nu}(\mathbf{x})$. The total Hamiltonian is then $\hat{H}=\hat{H}_0+\hat{H}^{CG}$, with $\hat{H}_0 =\int d^3\mathbf{x}\, T_{00}$ the free Hamiltonian for the field. Here $\hat{T}_{\mu\nu}$ is the free energy-momentum tensor operator. For a free complex scalar theory, it takes the standard form of $\hat{T}_{\mu\nu} = \partial_\mu \hat{\phi}^\dagger \partial_\nu \hat{\phi} + \partial_\nu \hat{\phi}^\dagger \partial_\mu \hat{\phi} - \eta_{\mu\nu}\partial_\rho \hat{\phi}^\dagger \partial^\rho \hat{\phi} - (m^2c^2/\hbar^2)\eta_{\mu\nu}\hat{\phi}^\dagger\hat{\phi}$. In the weak-field limit, the metric perturbation reduces to $h_{00}(x) \simeq -2\Phi_g(x)/c^2$, where $\Phi_g$ is the Newtonian potential of the source masses given by $\Phi_g(\mathbf{x},t) = -(G/c^2) \int d^3\mathbf{y} \langle \hat{T}_{00}(\mathbf{y},t) \rangle / |\mathbf{x} - \mathbf{y}|$. Substituting this into the Hamiltonian and keeping the dominant $00$ component gives
\begin{equation}
    \begin{split}
        \hat{H}^{\rm CG} \simeq -\frac{2}{c^2} \int d^3 \mathbf{x} \Phi_g (x) [\hat{\pi}^\dagger(\mathbf{x})\hat{\pi}(\mathbf{x}) - \frac{m^2 c^2}{2\hbar^2} \hat{\phi}^\dagger(\mathbf{x})\hat{\phi}(\mathbf{x})]
    \end{split}
\end{equation}
where $\hat{\pi} := \partial_0 \hat{\phi}$ is the conjugate momentum of the complex scalar field $\hat{\phi}$. In this semiclassical framework, the gravitational field acts as an external background rather than a dynamical quantum entity. Consequently, in an external-field expansion, gravitational lines represent insertions of the Newtonian background onto matter propagators rather than propagating gravitons. The relevant object is thus the dressed matter two-point function:
\begin{equation}
    G_g(x,y)=\frac{\langle 0|\mathcal{T}\hat{\phi}_{\rm{int}}(x)\hat{\phi}_{\rm{int}}^\dagger(y)\exp\{-i\int\hat{H}_{\mathrm{int}}^{\mathrm{CG}}\mathrm{d}t/\hbar\}|0\rangle}{\langle 0|\mathcal{T}\exp\{-i\int \hat{H}_{\mathrm{int}}^{\mathrm{CG}}\mathrm{d}t/\hbar \}|0\rangle},
\end{equation}
where the subscript `$\rm int$' represents interaction picture with respect to $\hat{H}_0$ in this letter. Perturbatively, this generates a Dyson series $G_g=G_0+G_0\Sigma_g G_0+\cdots$, representing the free matter propagator $G_0$ modified by vertex insertions $\Sigma_g$ from $\Phi_g$ as shown in Fig.~\ref{fig:placeholder}.

Let $f_A$ and $f_B$ be the mode functions of two localized positive-frequency modes $A,B$ centered in spatially separated regions. In the nonrelativistic limit relevant to massive-probe experiments, the creation operator is $\hat{a}_I^\dagger=\int \mathrm{d}^3\mathbf{x}\, f_I(\mathbf{x})\hat{\Psi}^\dagger(\mathbf{x})$. The exact mode-to-mode amplitude in the classical background is given by
\begin{equation}
    \begin{split}
        K_{A,B}(t) &:= \langle 0|\hat{a}_A \hat{U}_g (t)\hat{a}_B^\dagger|0\rangle\\
        &= \int \mathrm{d}^3\mathbf{x}\,\mathrm{d}^3\mathbf{y}\, f_A^*(\mathbf{x})G_g^{(+)}(\mathbf{x},t;\mathbf{y},0)f_B(\mathbf{y}),
    \end{split}
\end{equation}
where $\hat{U}_g = \mathcal{T}[e^{-i\int \mathrm{d} t \hat{H}^\text{CG}_\text{int}/\hbar}]e^{-i\hat{H}_0t/\hbar}$ is the evolution operator $\mathcal{T}e^{-i\int \hat{H}dt/\hbar}$ written in an interaction-picture form, and $G_g^{(+)}$ is the positive-frequency evolution kernel retaining only one-way propagation.

For two identical bosons occupying these modes, $|1_A,1_B\rangle = \hat{a}_A^\dagger \hat{a}_B^\dagger|0\rangle$, the survival amplitude decomposes into direct and exchange terms $\alpha_{AB} := \langle 1_A,1_B| \hat{U}_g |1_A,1_B\rangle = K_{A,A}K_{B,B} + K_{A,B}K_{B,A}$. Crucially, the crossed channel is dictated by $K_{A,B}K_{B,A}$, confirming it is strictly the off-diagonal part of the dressed matter propagator rather than an independent long-range interaction. 

This limitation directly extends to macroscopic objects. For $N$ identical bosons per mode, $|N_A,N_B\rangle = (1/N!)(\hat{a}_A^\dagger)^N(\hat{a}_B^\dagger)^N|0\rangle$, the leading exchange correction to the amplitude $\alpha_{AB}^{(N)} := \langle N_A,N_B| \hat{U}_g |N_A,N_B\rangle$ scales as
\begin{equation}
    \begin{split}
    \delta\alpha_{AB}^{(N)} ={}& N^2(K_{A,A}K_{B,B})^{N-1}K_{A,B}K_{B,A} \\
    &+ \mathcal{O}(K_{A,B}^2 K_{B,A}^2).
    \end{split}
\end{equation}
The combinatorial origin of this $N^2$ enhancement and the derivation of the exchange correction are worked out explicitly in Appendix~\ref{app:mode-exchange}. Thus, while macroscopic occupation provides an $N^2$ combinatorial enhancement, the physical channel remains fundamentally bounded by the off-diagonal propagation amplitude $K_{A,B}$. 

While macroscopic occupation provides a combinatorial enhancement $N^2$, it cannot remove the fundamental small parameter: the physical propagation amplitude $K_{A,B}$ between separated modes.

Applied to the four-branch geometry used in gravitational-entanglement proposals, with two spatial branches for single-constituent modes $|1L\rangle$, $|1R\rangle$ and $|2L\rangle$, $|2R\rangle$ for each object $1$ and $2$, the first-order crossed contribution is proportional to $K_{1R,2L}K_{2L,1R}$, while other cross terms are minor due to large spatial separation in the four-branch geometry\cite{aziz2025classical}.
Since the direct term $K_{1L/R,1L/R}K_{2L/R,2L/R}$ separates into the product of left and right amplitudes, it always factorize in the four-branch proposal, yielding no contribution to entanglement. Consequently, any entanglement attributed to this exchange mechanism is bounded strictly by the off-diagonal elements of the dressed propagator. The central question thus reduces to whether a classical background can render these off-diagonal elements appreciable between well-separated, physically localized systems. As we now demonstrate, it cannot: for bound matter they are exponentially localized, while for unbound finite-energy wave packets they reduce to ordinary overlap-suppressed propagation.

\begin{figure}[t]
    \centering
    \includegraphics[width=0.98\linewidth]{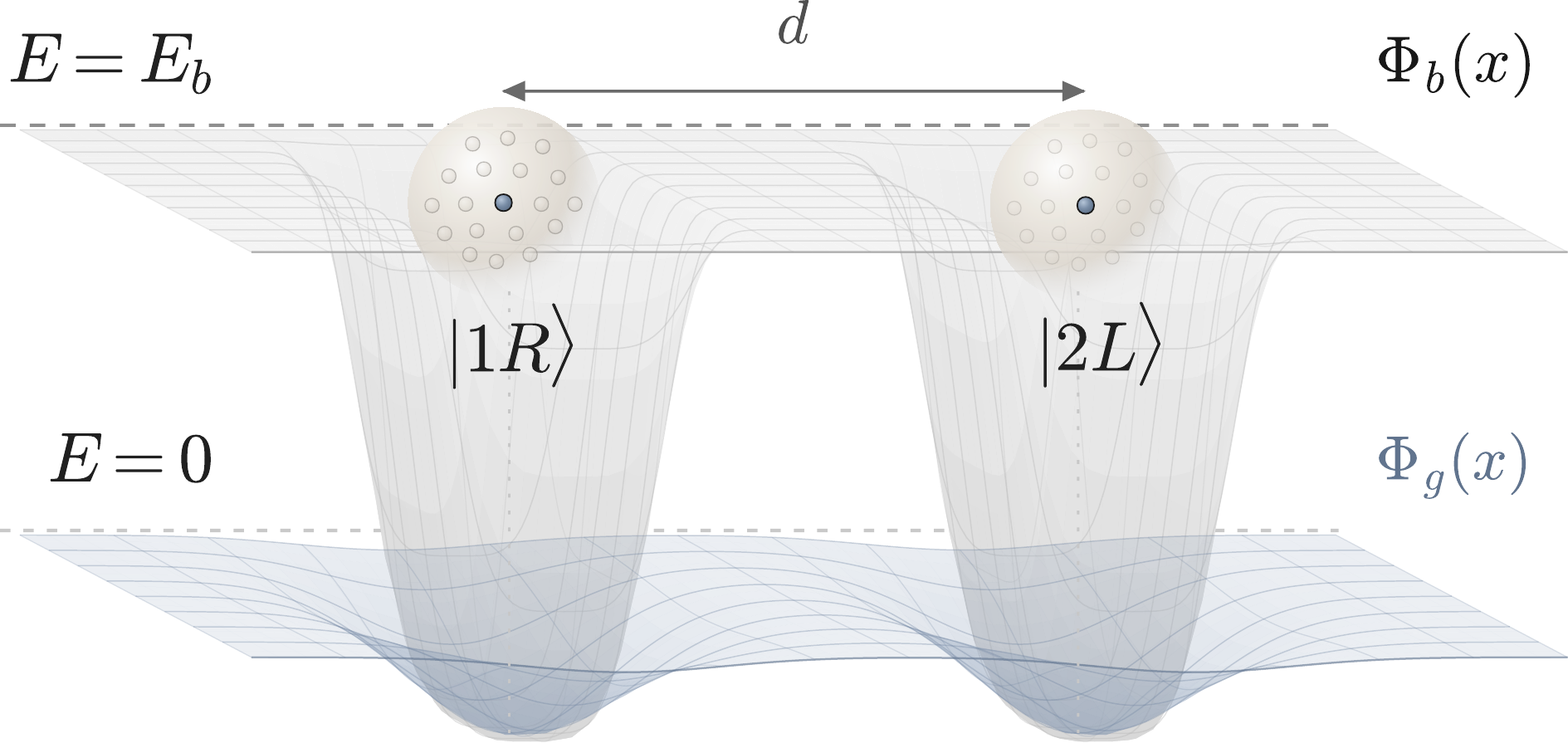}
    \caption{Schematic potential landscapes for two spatially localized matter configurations corresponding to the states $\lvert 1R\rangle$ and
    $\lvert 2L\rangle$, with center-to-center separation $d$. The binding potential $\Phi_b(x)$ (gray) is approximately
    constant inside each object and approaches $E_b$ outside, while the
    gravitational potential $\Phi_g(x)$ (blue) approaches zero asymptotically.
    Each translucent sphere consists of $N$ constituent atoms of mass $m$. the highlighted blue atom denotes a representative constituent moving in the mean field}
    \label{fig:bindingpotential}
\end{figure}

\emph{Localized matter: exponential suppression.---}
We first consider the case most directly relevant to solid masses, nanocrystals, and mechanical probes\cite{PhysRevLett.119.240401, PhysRevLett.119.240402, PhysRevA.102.062807, PhysRevResearch.3.023178, PhysRevD.102.106021, PhysRevD.110.024057, PhysRevD.107.106018, agafonova2026one}.  In such systems the microscopic constituents are not freely propagating scalar particles. They are localized by binding, lattice, or trapping potentials, and the low-energy states used to define the massive object are separated from states in which a constituent has escaped into the region between the two bodies. In such systems, each constituent experiences an effective single-particle binding potential $\Phi_b (\mathbf{x})$, which is approximately constant inside each localized object, with spatial scale $\sigma$, and rises to a barrier height $E_b$ outside. The total potential entering the dressed matter propagator is $\Phi (\mathbf{x}) = \Phi_g (\mathbf{x}) + \Phi_b (\mathbf{x})$, as shown in Fig.~\ref{fig:bindingpotential}. 
The binding potential modifies the propagator that enters the mode amplitude $K_{1R,2L}$ by shifting the pole of propagator to $\pm i \kappa$ with $\kappa=\sqrt{2mE_b}/\hbar$ and $E_b$ the binding energy, when $E_k\approx \hbar^2/2m\sigma^2\ll E_b \ll mc^2$. For sub-barrier propagation across spatial separation $d$ between the masses, this pole shift yields an exponential decay with anattenuation length $\ell = 1/\kappa = \hbar/\sqrt{2mE_b}$ of the off-diagonal mode amplitude. Up to algebraic prefactors, the $n$th-order contribution to the mode-to-mode amplitude is bounded by
\begin{equation}
    \begin{split}
        |K_{1R,2L}^{(n)}(t)|\lesssim \exp(-d/\ell),
    \end{split}
\end{equation}
where $d$ is the spatial separation between modes $|1R\rangle$ and $|2L\rangle$, while the finite-time dependence is contained in the prefactor. A detailed derivation of this bound is provided in Appendix~\ref{app:localized}. Equivalently, in the field-theoretic language, the free matter line appearing in the external-field expansion is replaced by the Green function of the localized matter Hamiltonian; for energies below the escape threshold, this Green function has no real on-shell momentum connecting the two bodies and decays exponentially with separation.

For typical microscopic constituents, $m\sim 10^{-27}\ \si{kg}$, and eV-scale binding energy $E_b$, this gives $\ell \sim 10\ \si{pm}$ on an atomic or subatomic length scale.  At the micrometer to millimeter separations $d$ relevant to massive-object gravitational-entanglement proposals, the amplitude is suppressed by a factor $\exp(-d/\ell) \sim  \exp(-10^{5\sim8})$. Even after the combinatorial enhancement of highly occupied modes, the leading exchange contribution scales as $N^2\exp(-2d/\ell)$, which is still negligible for any realistic macroscopic occupation number.

This conclusion does not depend on whether one describes the process as tunnelling, off-shell propagation, or virtual matter exchange.  The physical point is that a localized object is not described by a free matter field in the space between the bodies.  Once the binding responsible for the existence of the object is included, the matter-propagation channel required by the crossed diagram becomes short ranged.  Thus, for bound or condensed matter, the virtual-matter exchange mechanism cannot provide a long-range entangling channel mediated by a classical gravitational background.

\emph{Unbounded matter: restrictions from direct fusion} 
We next consider unbound matter, in which the effective constituents are not bound to one another into a localized composite object \cite{carney2021using,haine2021searching,howl2026gravitationally}. In practice, probing gravity with unbound matter remains experimentally challenging, since increasing the total mass requires coherent control of a large number of constituents while maintaining well-defined spatial modes and suppressing nongravitational interactions. Consequently, the achievable coherent mass is typically limited, making gravitational effects correspondingly weak. Nevertheless, we consider this regime for completeness. We denote by $m$ the mass of each effective constituent, whose internal degrees of freedom are not resolved, and by $M=Nm$ the total mass of $N$ such constituents. To isolate the unbound limit, we neglect nongravitational interactions between distinct constituents and assume that they evolve in a common classical gravitational potential sourced by the total matter distribution.

In this case the dressed propagator is the ordinary matter-evolution kernel in the classical gravitational field, and the crossed contribution remains controlled by the off-diagonal amplitudes \(K_{1R,2L}K_{2L,1R}\). To obtain this, we first calculate the interaction propagator with non-relativistic approximation $K^{(I)}_{\mathbf{y}\mathbf{x}}:=\langle \mathbf{y}|\mathcal{T} e^{-i\int \mathrm{d}t  \hat{H}^\text{CG}_\text{int}/\hbar}|\mathbf{x}\rangle$ of a single particle from one point $\mathbf{x}$ to another $\mathbf{y}$, where $\hat{H}^\text{CG}_\text{int}$ is the classical gravity potential generated by the two gaussian packets in the interaction picture as mentioned before. The non-relativistic approximation is proper, since all the energy scale involved in the analysis is far less than the particle mass. Note that the free Hamiltonian is shifted by a constant $mc^2$ in the non-relativistic approximation. A perturbative analysis in the weak-interaction regime gives the following bound for the $n$th-order contribution up to algebraic prefactor:
\begin{equation}
\left|K_{\mathbf{y}\mathbf{x}}^{(I,n)}(t)\right|
\lesssim
\left(
\frac{GNm^2 t}{d\hbar}
\right)^n
\exp\!\left[
-\frac{\sigma_t^2 m^2 d^2}{n t^2\hbar^2}
\right],
\label{eq:unbound-overlap}
\end{equation}
where $d=|\mathbf{x}-\mathbf{y}|$ is the propagation distance and $\sigma_t\sim\sqrt{\sigma_0^2+\hbar t/m}$ characterizes the wave-packet width at time $t$. Here the symbol $\lesssim$ absorbs numerical and algebraic prefactors arising from the ordered time integrations and momentum convolutions, none of which introduces an additional exponential dependence on $d$. A detailed derivation of this bound is provided in Appendix~\ref{app:unbound}. The remaining parameters have the same meanings as defined above.
 The off-diagonal amplitudes can be expressed by this quantity in the nonrelativisic limit as $K_{1R,2L}=\langle 1R|\mathcal{T} [e^{-i\int \mathrm{d}t \hat{H}^\text{CG}_\text{int} /\hbar}]e^{-i\hat{H}_0t/\hbar}|2L\rangle=\int \mathrm{d}\mathbf{x}\, \mathrm{d}\mathbf{y}\, \phi_{1R}^*(\mathbf{y},0)\phi_{2L}(\mathbf{x},t)K^{(I)}_{\mathbf{y}\mathbf{x}}$, with $\phi_I(\mathbf{x},t)$ the wave function of a particle initially of $I$-mode at time $t$ when no gravity exists. Notably, $\phi_I(\mathbf{x},0)$ is the mode function $f_I(\mathbf{x})$ of $I$-mode.

\begin{figure}[t]
    \centering
    \includegraphics[width=0.98\linewidth]{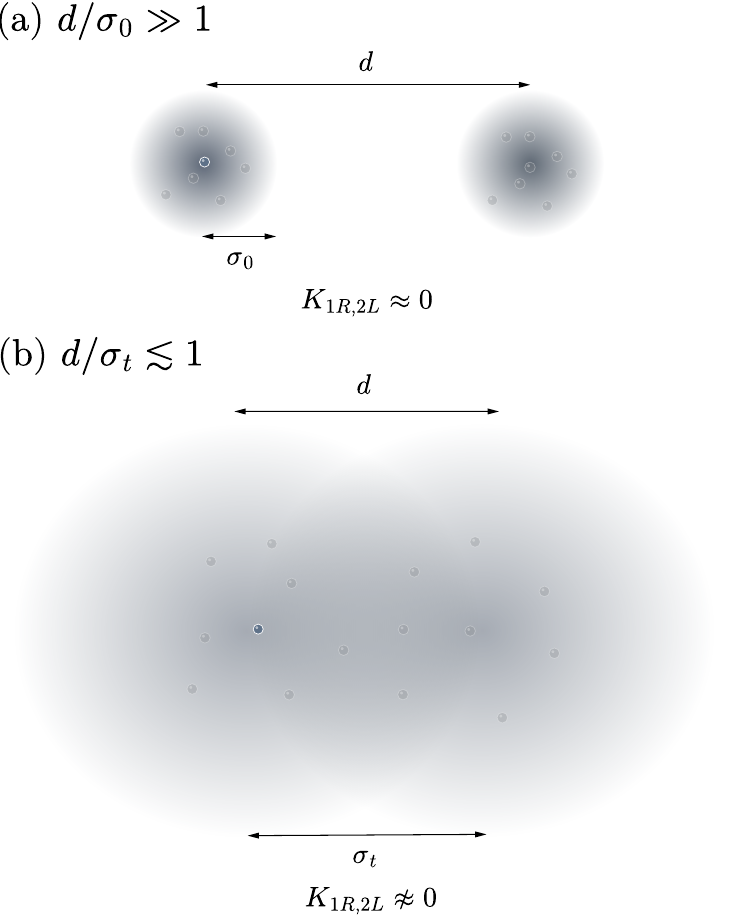}
    \caption{Direct fusion of unbound matter wave packets. (a) For $d/\sigma_0 \gg 1$, the modes $A$ and $B$ are spatially separated and off-diagonal matter propagation is negligible, such that $K_{1R,2L}\approx 0$. (b) Wave-packet spreading eventually yields $d/\sigma_t \lesssim 1$ and appreciable overlap, $K_{1R,2L}\not\approx 0$, at which point the left--right mode bipartition no longer defines two separated subsystems. The highlighted atom denotes a representative constituent.}
    \label{fig:overlap}
\end{figure}

Equation~\eqref{eq:unbound-overlap} leaves no regime in which this channel produces appreciable entanglement between separated systems, regardless of the parameter value. At large time scale, even without gravity interaction, particles initially in one mode will significantly overlap with the other mode after the evolution. Once they overlap appreciably, as quantified by $\int \mathrm{d}\mathbf{x}\,\mathrm{d}\mathbf{y}\,\phi_{1R}^*(\mathbf{x},0)\phi_{2L}(\mathbf{y},t)\delta^{(3)}(\mathbf{x}-\mathbf{y})$, the nominal left and right modes no longer constitute define independently addressable subsystems at such time scale. For identical particles, a left--right mode bipartition is then not physically well defined, and its entanglement cannot be identified with entanglement between two separated objects, as shown in Fig.~\ref{fig:overlap}. If $K^{(I)}_{\mathbf{y}\mathbf{x}}$ is suppressed, then there's no chance accumulating enough classical-gravity-induced-entanglements before the definition of entanglement breaks down. Namely, we require $\sigma_t^2m^2d^2 / t^2\hbar^2 \sim md^2 / t\hbar \lesssim 1$. Unfortunately, to maintain a well-defined entanglement, we need $d^2 / \sigma_t^2 \sim md^2 /\hbar t \gg 1$, contradicting with the previous requirement. Explicitly, this indicates that no significant classical-gravity-induced entanglement exists when the mode bipartition can be defined, even without considering the suppression from the prefactor of $K^{(I)}_{\mathbf{y}\mathbf{x}}$. The exponential suppression of $K^{(I)}_{\mathbf{y}\mathbf{x}}$ is relaxed at large $n$, but is again suppressed by the prefactor of Newton constant $G$, leading to no significant modification.

In conclusion, before
significant overlap the exchange channel is negligible, whereas after
significant overlap the required subsystem decomposition has ceased to
exist.

\emph{Origin of the apparent long-range contribution.—}
In a previous work \cite{aziz2025classical}, a contribution in the perturbative expansion shows a non-local propagation decaying as power law in distance instead of a mode overlap suppressed contribution. The apparent power-law contribution arises from taking the
infinite-time limit of a term in the Dyson expansion,
\begin{equation}
 \frac{\sin(uct/2)}{u}
 \xrightarrow[t\rightarrow\infty]{}
 \pi\delta(u),
 \label{eq:stationary-limit}
\end{equation}
which reduces the finite-time kernel to a stationary $1/r$ propagator. For large $r$, however, the integrand carries an oscillatory phase that prevents coherent accumulation until the time scale $t \sim m r^2/\hbar$ . Before this threshold, the oscillating factor leads to rapid cancellations, so the amplitude is strongly suppressed. The detailed finite-time restriction underlying this distributional limit is derived in Appendix~\ref{app:finite-time}.

At finite \(t\), every term in the dressed propagator is a convolution
of finite-time matter kernels. Transferring amplitude between the two
regions therefore requires the same wave-packet propagation that
controls their ordinary overlap, and the complete projected amplitude
obeys Eq.~\eqref{eq:unbound-overlap}. Taking \(t\rightarrow\infty\)
removes this finite-time localization information: any unbound
finite-energy packet eventually spreads across a fixed separation.
Consequently, the stationary \(1/r\) term is obtained only in a time regime
where the external packets can no longer be treated as fixed,
nonoverlapping modes.

The claimed long-range contribution thus results from combining two incompatible
assumptions: asymptotically long-time matter propagation and a persistent
bipartition into spatially separated modes. Before appreciable overlap,
the full crossed contribution is negligible; after appreciable overlap,
the two-mode entanglement used in the claim is no longer defined. A
discontinuous top-hat profile introduces additional ultraviolet
pathologies, while a self-consistent treatment also requires the
classical field to evolve with the spreading density, but neither
modification alters this central conclusion.

\textit{Conclusion}---We have shown that the proposed long-range entangling mechanism of classical gravity with crossing matter propagation cannot operate in the bound systems relevant to massive-object experiments under realistic conditions. Such experiments require binding, lattice, or trapping potentials to remain confined in stable and distinguishable spatial branches. Binding or trapping is therefore an essential component of the physical setup. Once this localization is included, constituent propagation between separated objects is sub-barrier and exponentially short ranged. Its microscopic attenuation length is far smaller than experimentally relevant separations, and the resulting suppression cannot be overcome by the enhancement associated with macroscopic occupation.

 For unbound wave packets, the contribution remains negligible while the modes represent distinct subsystems and becomes appreciable only after their spatial overlap has invalidated that interpretation. The apparent power-law behavior reported in \cite{aziz2025classical} results from combining an infinite-time approximation with an assumption that modes remain distinct. These assumptions are inherently incompatible. Once localization and wave-packet dynamics are treated consistently, the claimed long-range channel disappears. Our analysis therefore rules out this virtual-matter mechanism as a source of observable entanglement between spatially separated systems in a prescribed classical gravitational background. Entanglement as a sufficient evidence for quantum effects in gravity consequently remains intact.



\textit{Acknowledgment}--- Y. Liu acknowledges the support of Beijing Natural Science Foundation (Z240007), National Natural Science Foundation of China (No. 92365210 and No. 12374325), Young Elite Scientists Sponsorship Program by CAST (Grant No. 2023QNRC001), and Beijing Municipal Science and Technology Commission (Grant No. Z221100002722011).

\clearpage
\onecolumngrid
\appendix
\setcounter{secnumdepth}{3}
\allowdisplaybreaks[4]

\newcommand{\ii}{\mathrm{i}}
\newcommand{\dd}{\mathrm{d}}
\newcommand{\e}{\mathrm{e}}
\newcommand{\order}{\mathcal{O}}
\newcommand{\by}{\bm y}
\newcommand{\bz}{\bm z}
\newcommand{\br}{\bm r}
\newcommand{\bp}{\bm p}
\newcommand{\bq}{\bm q}
\newcommand{\bX}{\bm X}

The appendices comprise four sections, A--D. Appendix~A derives the mode-projected exchange amplitudes in Eqs.~(3) and (4) of the main text. Appendix~B derives the exponential bound in Eq.~(5) for localized constituents. Appendix~C derives the finite-time propagation bound underlying Eq.~(6) for unbound wave packets. Appendix~D establishes the finite-time restriction underlying the distributional limit in Eq.~(7). Symbols not defined in the main text are introduced upon first use.

\section{Detailed derivation of Eqs.~(3) and (4): mode projection and exchange}
\label{app:mode-exchange}

The main text defines the mode-to-mode amplitude
\begin{align}
K_{A,B}(t)
&:=
\bra{0}\hat a_A\hat U_g(t)\hat a_B^\dagger\ket{0}
\nonumber\\
&=
\int \dd^3x\,\dd^3y\,
f_A^*(\bx)\,
G_g^{(+)}(\bx,t;\by,0)\,
f_B(\by),
\label{eq:mainK}
\end{align}
where
\begin{equation}
\hat a_I^\dagger
=
\int \dd^3x\, f_I(\bx)\hat\Psi^\dagger(\bx)
\end{equation}
creates one particle in the localized positive-frequency mode $I$.
Let $\{\ket{\mu}\}$ be a complete orthonormal one-particle mode basis,
with wave functions $f_\mu(\bx)=\braket{\bx|\mu}$. In the
particle-number-conserving nonrelativistic sector, the positive-frequency kernel
propagates the creation field linearly, so that
\begin{align}
\hat U_g(t)\hat a_B^\dagger\hat U_g^\dagger(t)
&=
\int \dd^3y\,f_B(\by)\,
\hat U_g(t)\hat\Psi^\dagger(\by)\hat U_g^\dagger(t)
\nonumber\\
&=
\int \dd^3x\,\dd^3y\,
f_B(\by)\,
G_g^{(+)}(\bx,t;\by,0)\,
\hat\Psi^\dagger(\bx)
\nonumber\\
&=
\int \dd^3x\,\hat\Psi^\dagger(\bx)
\sum_\mu f_\mu(\bx)
\int \dd^3y\, f_\mu^*(\by)
\int \dd^3z\,
G_g^{(+)}(\by,t;\bz,0)f_B(\bz)
\nonumber\\
&=
\sum_\mu K_{\mu,B}(t)\hat a_\mu^\dagger .
\label{eq:creation-evolution}
\end{align}
Equation~\eqref{eq:creation-evolution} is simply Eq.~\eqref{eq:mainK} applied to
a complete output-mode basis; no additional propagation object is required.

For one identical boson in each of two orthogonal modes,
\begin{equation}
\ket{1_A,1_B}
=
\hat a_A^\dagger\hat a_B^\dagger\ket{0},
\qquad
\braket{A|B}=0,
\end{equation}
the survival amplitude is
\begin{align}
\alpha_{AB}(t)
&:=
\bra{1_A,1_B}\hat U_g(t)\ket{1_A,1_B}
\nonumber\\
&=
\bra0\hat a_B\hat a_A
\left[
\hat U_g(t)\hat a_A^\dagger\hat U_g^\dagger(t)
\right]
\left[
\hat U_g(t)\hat a_B^\dagger\hat U_g^\dagger(t)
\right]
\hat{U}_g(t)\ket0
\nonumber\\
&=
\sum_{\mu,\nu}
K_{\mu,A}(t)K_{\nu,B}(t)
\bra0\hat a_B\hat a_A
\hat a_\mu^\dagger\hat a_\nu^\dagger
\ket0
\nonumber\\
&=
\sum_{\mu,\nu}
K_{\mu,A}(t)K_{\nu,B}(t)
\left(
\delta_{A\mu}\delta_{B\nu}
+
\delta_{A\nu}\delta_{B\mu}
\right)
\nonumber\\
&=
K_{A,A}(t)K_{B,B}(t)
+
K_{A,B}(t)K_{B,A}(t),
\label{eq:two-boson}
\end{align}
where we have already used the invariance of vacuum under the evolution $|0\rangle=\hat{U}_g(t)|0\rangle$.The first term in Eq.~\eqref{eq:two-boson} is the direct contribution and the second
is the exchange contribution. Thus the crossed channel is entirely controlled by
the off-diagonal mode-projected propagators already defined in Eq.~(3) of the main text.

For $N$ identical bosons in each mode,
\begin{equation}
\ket{N_A,N_B}
=
\frac{(\hat a_A^\dagger)^N(\hat a_B^\dagger)^N}{N!}\ket0.
\end{equation}
Only the components returning to the $A,B$ sector survive the final projection, hence
\begin{align}
\alpha_{AB}^{(N)}(t)
&=
\frac{1}{(N!)^2}
\bra0\hat a_B^N\hat a_A^N
\left(
K_{A,A}\hat a_A^\dagger+K_{B,A}\hat a_B^\dagger
\right)^N
\left(
K_{A,B}\hat a_A^\dagger+K_{B,B}\hat a_B^\dagger
\right)^N
\ket0
\nonumber\\
&=
\frac{1}{(N!)^2}
\sum_{r,s=0}^{N}
\binom Nr\binom Ns
K_{A,A}^{N-r}K_{B,A}^{r}
K_{A,B}^{s}K_{B,B}^{N-s}
\bra0\hat a_B^N\hat a_A^N
(\hat a_A^\dagger)^{N-r+s}
(\hat a_B^\dagger)^{N+r-s}
\ket0.
\end{align}
The projection is nonzero only for $r=s\equiv k$, for which the last matrix element
equals $(N!)^2$. Therefore
\begin{equation}
\alpha_{AB}^{(N)}(t)
=
\sum_{k=0}^{N}
\binom Nk^2
\left(K_{A,A}K_{B,B}\right)^{N-k}
\left(K_{A,B}K_{B,A}\right)^k.
\label{eq:exact-N}
\end{equation}
Expanding Eq.~\eqref{eq:exact-N} in the small off-diagonal amplitude gives
\begin{align}
\alpha_{AB}^{(N)}
&=
\left(K_{A,A}K_{B,B}\right)^N
+
N^2
\left(K_{A,A}K_{B,B}\right)^{N-1}
K_{A,B}K_{B,A}
\nonumber\\
&\quad
+
\frac{N^2(N-1)^2}{4}
\left(K_{A,A}K_{B,B}\right)^{N-2}
\left(K_{A,B}K_{B,A}\right)^2
+\cdots,
\end{align}
and hence
\begin{equation}
\boxed{
\delta\alpha_{AB}^{(N)}
=
N^2
\left(K_{A,A}K_{B,B}\right)^{N-1}
K_{A,B}K_{B,A}
+
\order\!\left[
\left(K_{A,B}K_{B,A}\right)^2
\right]
}.
\label{eq:main4}
\end{equation}
This is Eq.~(4) of the main text.

The factorization of the direct contribution in the four-branch geometry follows
without introducing an additional entanglement measure. Indeed,
\begin{align}
\sum_{i,j=L,R}
\left(K_{1i,1i}\right)^N
\left(K_{2j,2j}\right)^N
\ket{1_i}\ket{2_j}
&=
\left[
\sum_{i=L,R}
\left(K_{1i,1i}\right)^N\ket{1_i}
\right]
\otimes
\left[
\sum_{j=L,R}
\left(K_{2j,2j}\right)^N\ket{2_j}
\right].
\end{align}
It is therefore the nonfactorizing crossed term
$K_{1i,2j}K_{2j,1i}$, not the direct propagation, that can contribute to branch
entanglement.

\section{Detailed derivation of Eq.~(5): localized constituents and exponential attenuation}
\label{app:localized}

For a physically localized constituent, the Hamiltonian responsible for localization
must be included in the unperturbed matter evolution.  We keep the same
classical-background convention as in the main text: $\Phi_b(\bx)$ and
$\Phi_g(\bx,t)$ are c-number functions of the spatial coordinate.  Their action
on the one-particle matter Hilbert space is represented by
\begin{align}
\hat V_b
&:=
\int\dd^3x\,
m\Phi_b(\bx)\,
\ket{\bx}\bra{\bx},
\\
\hat V_g(t)
&:=
\int\dd^3x\,
m\Phi_g(\bx,t)\,
\ket{\bx}\bra{\bx},
\end{align}
so that
\begin{align}
\bra{\bx}\hat V_b\ket{\by}
&=
m\Phi_b(\bx)\delta^{(3)}(\bx-\by),
\\
\bra{\bx}\hat V_g(t)\ket{\by}
&=
m\Phi_g(\bx,t)\delta^{(3)}(\bx-\by).
\end{align}
Thus the gravitational field remains a prescribed classical background; the hat
on $\hat V_g$ only denotes the operator by which this c-number potential acts on
the quantum matter Hilbert space.  We write
\begin{equation}
\hat H(t)
=
\hat H_b+\hat V_g(t),
\qquad
\hat H_b
=\frac{\hat{\bm p}^{\,2}}{2m}
+
\hat V_b,
\label{eq:Hbound}
\end{equation}
where we have already discarded the rest energy $mc^2$.

Let $\ket{1R}$ and $\ket{2L}$ denote the single-constituent branch modes used in
the main text, with
\begin{equation}
f_{1R}(\bx)=\braket{\bx|1R},
\qquad
f_{2L}(\bx)=\braket{\bx|2L}.
\end{equation}
Using $\hat H_b$ as the unperturbed Hamiltonian, the Furry-picture decomposition is
\begin{align}
\hat U_g(t)
&=
\e^{-\ii\hat H_b t/\hbar}\hat U_g^{F}(t),
\\
\hat U_g^{F}(t)
&=
\mathcal T
\exp\!\left[
-\frac{\ii}{\hbar}
\int_0^t\dd t_1\,
\hat V_g^{F}(t_1)
\right],
\\
\hat V_g^F(t_1)
&=
\e^{+\ii\hat H_b t_1/\hbar}
\hat V_g(t_1)
\e^{-\ii\hat H_b t_1/\hbar}.
\end{align}
The $n$th-order contribution to the off-diagonal one-constituent amplitude appearing
in Eq.~(5) of the main text is therefore
\begin{align}
K_{1R,2L}^{(n)}(t)
&:=
\left(-\frac{\ii}{\hbar}\right)^n
\int_{0<t_n<\cdots<t_1<t}
\prod_{a=1}^{n}\dd t_a
\nonumber\\
&\quad\times\bra{1R}
\e^{-\ii\hat H_b(t-t_1)/\hbar}
\hat V_g(t_1)
\e^{-\ii\hat H_b(t_1-t_2)/\hbar}
\hat V_g(t_2)
\cdots
\hat V_g(t_n)
\e^{-\ii\hat H_b t_n/\hbar}
\ket{2L},
\quad n\ge1.
\label{eq:K12-bound-finite}
\end{align}
Equation~\eqref{eq:K12-bound-finite} shows explicitly that every propagation
interval between neighboring gravitational insertions is generated by the localized
matter Hamiltonian $\hat H_b$, rather than by the free Hamiltonian.

To expose the corresponding spatial attenuation, define the energy resolvent operator
\begin{equation}
\hat G_b^{(+)}(E)
:=
\frac{1}{E-\hat H_b+\ii0},
\label{eq:Gb-operator}
\end{equation}
and its coordinate-space kernel
\begin{equation}
G_b^{(+)}(\bx,\by;E)
:=
\bra{\bx}\hat G_b^{(+)}(E)\ket{\by}.
\label{eq:Gb-kernel-def}
\end{equation}
The resolvent is related to the $\hat H_b$-generated time evolution by
\begin{equation}
\hat G_b^{(+)}(E)
=
-\frac{\ii}{\hbar}
\int_0^\infty\dd\tau\,
\e^{+\ii(E+\ii0)\tau/\hbar}
\e^{-\ii\hat H_b\tau/\hbar}.
\label{eq:resolvent-time}
\end{equation}
Hence, in the stationary reduction, the time intervals between neighboring
$\hat V_g$ insertions in Eq.~\eqref{eq:K12-bound-finite} become insertions of
$\hat G_b^{(+)}(E)$.

Suppose the two localized branch modes are narrow in energy around $E$ under
$\hat H_b$,
\begin{equation}
\hat H_b\ket{1R}\simeq E\ket{1R},
\qquad
\hat H_b\ket{2L}\simeq E\ket{2L}.
\end{equation}
For the quasistatic background relevant to the bound-matter estimate,
$\Phi_g(\bx,t)\simeq\Phi_g(\bx)$, the second-order term becomes
\begin{align}
K_{1R,2L}^{(2)}(t)
&=
-\frac{\e^{-\ii Et/\hbar}}{\hbar^2}
\int\dd E'\,
\bra{1R}\hat V_g\ket{E'}
\bra{E'}\hat V_g\ket{2L}
\int_0^t\dd t_1
\int_0^{t_1}\dd t_2\,
\e^{-\ii(E'-E)(t_1-t_2)/\hbar}
\nonumber\\
&\simeq
-\frac{\ii t}{\hbar}
\e^{-\ii Et/\hbar}
\bra{1R}
\hat V_g\hat G_b^{(+)}(E)\hat V_g
\ket{2L}.
\label{eq:K12-second-resolvent}
\end{align}
Repeating the same stationary reduction between every neighboring pair of
insertions gives
\begin{equation}
K_{1R,2L}^{(n)}(t)
\simeq
-\frac{\ii t}{\hbar}
\e^{-\ii Et/\hbar}
\bra{1R}
\hat V_g
\left[
\hat G_b^{(+)}(E)\hat V_g
\right]^{n-1}
\ket{2L},
\qquad n\ge1.
\label{eq:K12-resolvent}
\end{equation}
The finite-time dependence is therefore contained in the prefactor, whereas the
long-distance behavior is determined by the spatial kernel of
$\hat G_b^{(+)}(E)$.

In the barrier region, the binding potential energy is approximately constant,
\begin{equation}
m\Phi_b(\bx)\simeq E_b,
\end{equation}
and a localized constituent has $E<E_b$.  Equation~\eqref{eq:Gb-kernel-def} gives
\begin{align}
|G_b^{(+)}(\bx,\by;E)|
&\lesssim|
\int\frac{\dd^3q}{(2\pi)^3}
\frac{\e^{\ii\bq\cdot(\bx-\by)}}
{E-E_b-\hbar^2q^2/(2m)+\ii0}|.
\end{align}
This is more explicit in the path integral formalism, where every path linking $\bx,\by$ from different matter has to goes through such barrier region, and thus the magnitude of the Green function is obviously controlled by amplitude from such process. 
Defining
\begin{equation}
\kappa(E)
=
\frac{\sqrt{2m(E_b-E)}}{\hbar},
\qquad
E-E_b
=
-\frac{\hbar^2\kappa^2}{2m},
\end{equation}
and setting $r:=|\bx-\by|$, one obtains
\begin{align}
|G_b^{(+)}(\bx,\by;E)|
&\lesssim
|-\frac{2m}{\hbar^2}
\int\frac{\dd^3q}{(2\pi)^3}
\frac{\e^{\ii\bq\cdot(\bx-\by)}}{q^2+\kappa^2}|
\nonumber\\
&=
\frac{m}{\pi^2\hbar^2r}
\int_0^\infty\dd q\,
\frac{q\sin(qr)}{q^2+\kappa^2}
\nonumber\\
&=
\frac{m}{2\pi\hbar^2}
\frac{\e^{-\kappa r}}{r}.
\label{eq:Yukawa}
\end{align}
Thus the matter propagation between gravitational insertions is evanescent.

In the regime used in the main text,
\begin{equation}
E\simeq E_k
\approx
\frac{\hbar^2}{2m\sigma^2}
\ll E_b
\ll mc^2,
\end{equation}
so that
\begin{equation}
\kappa
\simeq
\frac{\sqrt{2mE_b}}{\hbar},
\qquad
\ell
:=
\kappa^{-1}
=
\frac{\hbar}{\sqrt{2mE_b}}.
\label{eq:ell}
\end{equation}

We now derive the fixed-order spatial bound explicitly.  Let
\begin{equation}
\Phi_*
:=
\sup_{\bx}|\Phi_g(\bx)|,
\qquad
\|f_I\|_1
:=
\int\dd^3x\,|f_I(\bx)|.
\end{equation}
For $n\ge2$, insert $n$ coordinate resolutions of the identity into
Eq.~\eqref{eq:K12-resolvent}:
\begin{align}
K_{1R,2L}^{(n)}(t)
&\simeq
-\frac{\ii t}{\hbar}\e^{-\ii Et/\hbar}
\int\prod_{a=1}^{n}\dd^3x_a\,
\bra{1R}\hat V_g\ket{\bx_1}
\prod_{a=1}^{n-1}
\left[
\bra{\bx_a}\hat G_b^{(+)}(E)\hat V_g\ket{\bx_{a+1}}
\right]
\braket{\bx_n|2L}.
\label{eq:K12-coordinate-insertion}
\end{align}
The endpoint factors are
\begin{equation}
\bra{1R}\hat V_g\ket{\bx_1}
=
f_{1R}^*(\bx_1)m\Phi_g(\bx_1),
\qquad
\braket{\bx_n|2L}
=
f_{2L}(\bx_n),
\end{equation}
while each internal factor is
\begin{align}
\bra{\bx_a}\hat G_b^{(+)}(E)\hat V_g\ket{\bx_{a+1}}
&=
\int\dd^3z_a\,
G_b^{(+)}(\bx_a,\bz_a;E)
m\Phi_g(\bz_a)
\delta^{(3)}(\bz_a-\bx_{a+1})
\nonumber\\
&=
mG_b^{(+)}(\bx_a,\bx_{a+1};E)
\Phi_g(\bx_{a+1}).
\end{align}
Therefore, before taking the modulus,
\begin{align}
K_{1R,2L}^{(n)}(t)
&\simeq
-\frac{\ii t}{\hbar}
\e^{-\ii Et/\hbar}
m^n
\int\prod_{a=1}^{n}\dd^3x_a\,
f_{1R}^*(\bx_1)\Phi_g(\bx_1) 
\prod_{a=1}^{n-1}
\left[
G_b^{(+)}(\bx_a,\bx_{a+1};E)
\Phi_g(\bx_{a+1})
\right]
f_{2L}(\bx_n).
\label{eq:K12-scalar-G}
\end{align}
Substituting Eq.~\eqref{eq:Yukawa}, with
$r_{a,a+1}:=|\bx_a-\bx_{a+1}|$, gives
\begin{align}
|K_{1R,2L}^{(n)}(t)|
&\le
\frac{t\,m^{2n-1}}
{(2\pi)^{n-1}\hbar^{2n-1}}
\int\prod_{a=1}^{n}\dd^3x_a\,
|f_{1R}^*(\bx_1)||\Phi_g(\bx_1)|
\prod_{a=1}^{n-1}
\left[
\frac{\e^{-\kappa r_{a,a+1}}}{r_{a,a+1}}
|\Phi_g(\bx_{a+1})|
\right]
|f_{2L}(\bx_n)|.
\label{eq:K12-scalar-Yukawa}
\end{align}
Taking the modulus and using $|\Phi_g(\bx)|\le\Phi_*$,
\begin{align}
|K_{1R,2L}^{(n)}(t)|
&\le
\frac{t\,m^{2n-1}\Phi_*^n}
{(2\pi)^{n-1}\hbar^{2n-1}}
\int\prod_{a=1}^{n}\dd^3x_a\,
|f_{1R}(\bx_1)|\,|f_{2L}(\bx_n)|
\prod_{a=1}^{n-1}
\frac{\e^{-\kappa r_{a,a+1}}}{r_{a,a+1}} .
\label{eq:K12-after-modulus}
\end{align}
Introduce the Yukawa kernel
\begin{equation}
Y_\kappa(\br)
:=
\frac{\e^{-\kappa|\br|}}{4\pi|\br|}
=
\int\frac{\dd^3q}{(2\pi)^3}
\frac{\e^{\ii\bq\cdot\br}}{q^2+\kappa^2}.
\end{equation}
The $n-2$ intermediate coordinate integrations form the $(n-1)$-fold convolution
\begin{align}
I_L(r)
&:=
Y_\kappa^{*L}(r)
=
\int\frac{\dd^3q}{(2\pi)^3}
\frac{\e^{\ii\bq\cdot\br}}{(q^2+\kappa^2)^L}
\nonumber\\
&=
\frac{1}
{(2\pi)^{3/2}2^{L-1}\Gamma(L)}
\left(\frac{r}{\kappa}\right)^{L-3/2}
K_{L-3/2}(\kappa r),
\end{align}
where $K_\nu$ is the modified Bessel function of the second kind.  If $d$ denotes
the spatial separation between the two localized branch modes, the endpoint
separation is bounded from below by $d$ up to width-scale corrections.  Hence, in
the separated regime,
\begin{align}
|K_{1R,2L}^{(n)}(t)|
&\le
\frac{
2t\,m^{2n-1}\Phi_*^n
\|f_{1R}\|_1\|f_{2L}\|_1
}
{
(2\pi)^{3/2}\hbar^{2n-1}\Gamma(n-1)
}
\left(\frac{d}{\kappa}\right)^{n-5/2}
K_{n-5/2}(\kappa d),
\qquad n\ge2.
\label{eq:K12-Bessel-bound}
\end{align}
For $\kappa d\gg1$,
\begin{equation}
K_\nu(z)
=
\sqrt{\frac{\pi}{2z}}
\e^{-z}
\left[
1+\order(z^{-1})
\right],
\end{equation}
so that
\begin{align}
|K_{1R,2L}^{(n)}(t)|
&\lesssim
\frac{
t\,m^{2n-1}\Phi_*^n
\|f_{1R}\|_1\|f_{2L}\|_1
}
{
2\pi\hbar^{2n-1}(n-2)!
}
\frac{d^{\,n-3}}{\kappa^{\,n-2}}
\e^{-\kappa d},
\qquad n\ge2.
\label{eq:K12-fixed-order}
\end{align}
Thus, at every fixed perturbative order, the dependence on $d$ outside the Yukawa
factor is algebraic.  The first-order term,
\begin{equation}
K_{1R,2L}^{(1)}(t)
\simeq
-\frac{\ii t}{\hbar}
\e^{-\ii Et/\hbar}
\int\dd^3x\,
f_{1R}^*(\bx)\,
m\Phi_g(\bx)\,
f_{2L}(\bx),
\end{equation}
vanishes for strictly disjoint compact supports and is overlap-suppressed for
realistic localized modes.  Therefore the $n$th-order contribution quoted in the
main text retains the exponential spatial suppression
\begin{equation}
\boxed{
|K_{1R,2L}^{(n)}(t)|
\lesssim
\e^{-\kappa d}
=
\e^{-d/\ell},
}
\label{eq:main5}
\end{equation}
where the finite-time dependence is contained in the algebraic prefactor.

The stationary reduction above is controlled when the finite-time energy window
resolves the binding gap.  Since
\begin{align}
A_t(u)
&=
\int_0^t\dd\tau\,\e^{\ii u\tau}
=
\e^{\ii ut/2}
\frac{2\sin(ut/2)}{u},
\end{align}
the energy resolution is $\Delta E\sim\hbar/t$, and the stationary form requires
\begin{equation}
\frac{\hbar}{t}
\ll
E_b-E
\qquad\Longleftrightarrow\qquad
t
\gg
\frac{\hbar}{E_b-E}.
\label{eq:bound-time}
\end{equation}
At shorter times one retains the exact finite-time expression
Eq.~\eqref{eq:K12-bound-finite}; the propagation intervals are still generated by
the localized Hamiltonian $\hat H_b$.

\section{Detailed derivation of Eq.~(6): unbound finite-energy wave packets}
\label{app:unbound}

We now remove the binding potential and take
\begin{equation}
\hat H_0
=
\frac{\hat{\bm p}^{\,2}}{2m}.
\end{equation}
The interaction-picture evolution generated by the prescribed classical
gravitational background is
\begin{equation}
\hat U_I(t)
=
\mathcal T
\exp\!\left[
-\frac{\ii}{\hbar}
\int_0^t\dd t_1\,
\hat H_{\rm int}^{\rm CG}(t_1)
\right].
\end{equation}
As in the main text, define the single-particle interaction kernel
\begin{equation}
K^{(I)}_{yx}(t)
:=
\braket{\by|\hat U_I(t)|\bx}
=
\sum_{n=0}^{\infty}
K^{(I,n)}_{yx}(t),
\end{equation}
with
\begin{align}
K^{(I,n)}_{yx}(t)
&=
\left(-\frac{\ii}{\hbar}\right)^n
\int_{0<t_1<\cdots<t_n<t}
\prod_{j=1}^{n}\dd t_j\,
\bra{\by}
\hat H_{\rm int}^{\rm CG}(t_n)\cdots
\hat H_{\rm int}^{\rm CG}(t_1)
\ket{\bx}.
\label{eq:Dyson-unbound}
\end{align}

Let the two unbound Gaussian source distributions have total mass $M=Nm$, centers
$\bX_1,\bX_2$, and width $\sigma_t$:
\begin{equation}
\rho_I(\bx,t)
=
\frac{M}{(2\pi\sigma_t^2)^{3/2}}
\exp\!\left[
-\frac{|\bx-\bX_I|^2}{2\sigma_t^2}
\right],
\qquad
I=1,2.
\label{eq:rho}
\end{equation}
With the Fourier convention
\begin{equation}
\widetilde\rho_I(\bq,t)
=
\int\dd^3x\,
\e^{-\ii\bq\cdot\bx/\hbar}
\rho_I(\bx,t),
\end{equation}
one has
\begin{equation}
\widetilde\rho_I(\bq,t)
=
M
\exp\!\left[
-\frac{\sigma_t^2q^2}{2\hbar^2}
\right]
\e^{-\ii\bq\cdot\bX_I/\hbar}.
\end{equation}
The classical gravitational potential acts on one constituent through
\begin{equation}
\hat V_g(t)
=
\int\dd^3x\,
m\Phi_g(\bx,t)\ket{\bx}\bra{\bx},
\end{equation}
with
\begin{equation}
\bra{\bx}\hat V_g(t)\ket{\by}
=
m\Phi_g(\bx,t)\delta^{(3)}(\bx-\by).
\end{equation}
Poisson's equation gives the Fourier transform of the corresponding c-number
potential energy,
\begin{align}
\widetilde V_g(\bq,t)
&=
-\frac{4\pi GmM\hbar^2}{q^2}
\exp\!\left[
-\frac{\sigma_t^2q^2}{2\hbar^2}
\right]
\left(
\e^{-\ii\bq\cdot\bX_1/\hbar}
+
\e^{-\ii\bq\cdot\bX_2/\hbar}
\right),
\label{eq:Vq}
\\
|\widetilde V_g(\bq,t)|
&\le
\frac{8\pi GmM\hbar^2}{q^2}
\exp\!\left[
-\frac{\sigma_t^2q^2}{2\hbar^2}
\right].
\label{eq:Vq-bound}
\end{align}

Insert momentum resolutions,
\begin{equation}
\braket{\bx|\bp}
=
\frac{1}{(2\pi\hbar)^{3/2}}
\e^{\ii\bp\cdot\bx/\hbar},
\qquad
\mathbf 1
=
\int\dd^3p\,\ket{\bp}\bra{\bp},
\end{equation}
into Eq.~\eqref{eq:Dyson-unbound}.  Here $\bp_j$ denotes a momentum integration
variable.  Defining the momentum transfer at the $j$th insertion by
\begin{equation}
\bq_j
:=
\bp_j-\bp_{j-1},
\end{equation}
the initial momentum $\bp_0$ appears linearly in the total phase and can be
integrated exactly:
\begin{align}
\int\frac{\dd^3p_0}{(2\pi\hbar)^3}
\exp\!\left[
\frac{\ii}{\hbar}
\bp_0\cdot
\left(
\by-\bx+
\frac{1}{m}\sum_{j=1}^{n}t_j\bq_j
\right)
\right]
&=
\delta^{(3)}
\left(
\by-\bx+
\frac{1}{m}\sum_{j=1}^{n}t_j\bq_j
\right).
\end{align}
The exact $n$th-order interaction kernel is therefore
\begin{align}
K^{(I,n)}_{yx}(t)
&=
\left(-\frac{\ii}{\hbar}\right)^n
\int_{0<t_1<\cdots<t_n<t}
\prod_{j=1}^{n}\dd t_j
\int
\prod_{j=1}^{n}
\frac{\dd^3q_j}{(2\pi\hbar)^3}
\delta^{(3)}
\left(
\by-\bx+
\frac{1}{m}\sum_{j=1}^{n}t_j\bq_j
\right)
\e^{\ii\varphi_n/\hbar}
\prod_{j=1}^{n}
\widetilde V_g(\bq_j,t_j),
\label{eq:exact-q}
\end{align}
where $\varphi_n$ is a real phase generated by the free kinetic evolution and
endpoint phases.  Since $|\e^{\ii\varphi_n/\hbar}|=1$, this phase does not affect
the modulus bound.

Let $d:=|\bx-\by|$ in agreement with the main text.  The delta function in Eq.~\eqref{eq:exact-q}
imposes
\begin{equation}
\sum_{j=1}^{n}t_j\bq_j
=
-m(\by-\bx).
\end{equation}
Consequently,
\begin{align}
m^2d^2
&=
\left|
\sum_{j=1}^{n}
\frac{t_j}{\sigma_{t_j}}
\left(
\sigma_{t_j}\bq_j
\right)
\right|^2
\le
\left(
\sum_{j=1}^{n}
\frac{t_j^2}{\sigma_{t_j}^2}
\right)
\left(
\sum_{j=1}^{n}
\sigma_{t_j}^2q_j^2
\right),
\end{align}
which gives
\begin{equation}
\sum_{j=1}^{n}
\sigma_{t_j}^2q_j^2
\ge
\frac{m^2d^2}
{\displaystyle
\sum_{j=1}^{n}t_j^2/\sigma_{t_j}^2}.
\label{eq:weighted-CS}
\end{equation}
The Gaussian factors associated with the source profiles therefore obey
\begin{equation}
\prod_{j=1}^{n}
\exp\!\left[
-\frac{\sigma_{t_j}^2q_j^2}{2\hbar^2}
\right]
\le
\exp\!\left[
-\frac{m^2d^2}
{2\hbar^2
\displaystyle\sum_{j=1}^{n}t_j^2/\sigma_{t_j}^2}
\right].
\label{eq:Gaussian-exact}
\end{equation}
This is the finite-time propagation suppression before any stationary
large-$t$ replacement.

For the compact scaling form quoted in Eq.~(6) of the main text, the $n$ insertion
times are of order $t$ and the source widths are characterized by the common scale
$\sigma_t$, so that parametrically
\begin{equation}
\sum_{j=1}^{n}
\frac{t_j^2}{\sigma_{t_j}^2}
=
\order\!\left(
\frac{nt^2}{\sigma_t^2}
\right).
\end{equation}
At fixed perturbative order, the remaining $q_j^{-2}$ Coulomb kernels, the ordered
time domain, and the dimension-carrying factors from the momentum convolutions
contribute only numerical and algebraic prefactors.  Using $M=Nm$ and retaining
the weak-gravity scale and the Gaussian propagation factor displayed in the main
text, Eq.~\eqref{eq:exact-q} therefore gives
\begin{equation}
\boxed{
\left|K^{(I,n)}_{yx}(t)\right|
\lesssim
\left(
\frac{GNm^2t}{d\hbar}
\right)^n
\exp\!\left[
-\frac{\sigma_t^2m^2d^2}
{nt^2\hbar^2}
\right].
}
\label{eq:main6}
\end{equation}
Here $\lesssim$ absorbs the numerical and algebraic prefactors from the ordered
time integrations and momentum convolutions, including order-unity constants
associated with the precise Gaussian-width convention.  No additional
distance-dependent exponential is introduced by these prefactors.  This is the form used in Eq.~(6) of the main text.

For a freely spreading minimum-uncertainty Gaussian with no initial
position-momentum correlation,
\begin{equation}
\sigma_t^2
=
\sigma_0^2
+
\left(
\frac{\hbar t}{2m\sigma_0}
\right)^2
\ge
\frac{\hbar t}{m}.
\label{eq:sigmat}
\end{equation}
The main text uses the equivalent scaling estimate
$\sigma_t\sim\sqrt{\sigma_0^2+\hbar t/m}$, since only the parametric spreading
scale is needed.  Equation~\eqref{eq:Gaussian-exact} then implies
\begin{align}
\sum_{j=1}^{n}
\frac{t_j^2}{\sigma_{t_j}^2}
&\le
\frac{m}{\hbar}
\sum_{j=1}^{n}t_j
\le
\frac{nmt}{\hbar},
\end{align}
and hence, keeping the explicit order-unity coefficient in this intermediate
bound,
\begin{equation}
\left|K^{(I,n)}_{yx}(t)\right|
\le
(\text{algebraic and coupling prefactors})
\exp\!\left[
-\frac{md^2}{2n\hbar t}
\right].
\label{eq:rigorous-fixed-order}
\end{equation}
Thus, for every fixed $n=O(1)$, the finite-time interaction kernel is
exponentially suppressed while $md^2/(\hbar t)\gg1$.

Finally, the off-diagonal branch amplitude used in the main text is obtained by
projecting the interaction kernel onto the freely evolved packet modes.  For the
example written there,
\begin{equation}
K_{1R,2L}(t)
=
\int\dd^3x\,\dd^3y\,
\phi_{1R}^*(\by,0)\,
\phi_{2L}(\bx,t)\,
K^{(I)}_{yx}(t),
\label{eq:mode-projection-unbound}
\end{equation}
where $\phi_I(\bx,t)$ is the freely evolved wave function of a particle initially
in mode $I$, and $\phi_I(\bx,0)=f_I(\bx)$.  As long as the packets remain narrow
compared with their separation, this projection changes $d$ only by width-scale
corrections and does not remove the finite-time propagation suppression.

\section{Detailed derivation of Eq.~(7): finite-time threshold limit and mode distinguishability}
\label{app:finite-time}

The long-range stationary term discussed in the main text and derived in Ref.~\cite{aziz2025classical} originates from
\begin{align}
A_t
&\propto
\int_t\dd^4x\,\int_t\dd^4z\,\int \frac{d^{4}k}{(2\pi)^{4}}\,
\frac{1}{k^{2}+m^{2}c^{2}/\hbar^{2}}\,
e^{ik\cdot(x-z)}\,
f_{2j}(\bx)
f_{1i}(\bz)
\Phi(\bx)
\Phi(\bz)
e^{imc(z^{0}-x^{0})/\hbar}
\,.
\end{align}
The time dependant kernel, set $\gamma= mc/\hbar$:
\begin{align}
I_t(\bx,\bz)
&=
\int_0^{ct} \dd x^0\,\int_0^{ct}\dd z^0\,\int \frac{d^{4}k}{(2\pi)^{4}}\,
\frac{1}{k^{2}+m^{2}c^{2}/\hbar^{2}}\,
e^{ik\cdot(x-z)}\,e^{i\gamma(z^{0}-x^{0})}\,.
\end{align}
Integrate out spatial components of $k$ to get:
\begin{align}
I_t
&=
\int \frac{dk^0}{2\pi}\,
f(k^0)
\left[
\int_0^{ct} dx^0\,
e^{\,i(k^0-\gamma)x^0}
\right]
\left[
\int_0^{ct} dz^0\,
e^{-i(k^0-\gamma)z^0}
\right].
\end{align}
where:
\begin{equation}
    f(k^0)=\frac{1}{4\pi r}
\Bigg[
\theta\!\left((k^0)^2-\gamma^2\right)
\left(
e^{-r\sqrt{\gamma^2-(k^0)^2}}-1
\right)
+\theta\!\left(\gamma^2-(k^0)^2\right)
\left(
\cos\!\left(r\sqrt{(k^0)^2-\gamma^2}\right)-1
\right)
+1
\Bigg]
\end{equation}
We now perform the time integral over $x^0$,
\begin{align}
I_t
&=
\int_0^{ct}\dd z^0
\int\frac{\dd k^0}{2\pi}\,
f(k^0)\,
e^{-i(k^0-\gamma)z^0}
\,2e^{ict(k^0-\gamma)/2}
\frac{\sin\!\left[ct(k^0-\gamma)/2\right]}{k^0-\gamma}.
\end{align}
If we take the large-interaction-time approximation as in Ref.~\cite{aziz2025classical}, using
\begin{equation}
    \boxed{\frac{\sin(ctu/2)}{u}\xrightarrow{t\to \infty}\pi\delta(u)}
\end{equation}
Writing $u=k^0-\gamma$ and using, for large interaction time this becomes
\begin{align}
I_t
&\approx
\int_0^{ct}\dd z^0
\int \dd u\,
e^{ictu/2}e^{-iuz^0}
\delta(u)\,f(u+\gamma)
=
\int_0^{ct}\dd z^0\,f(\gamma)
=
\frac{ct}{4\pi r},
\end{align}
where $r=|\bx-\bz|$ and
$
f(\gamma)=1/(4\pi r)
$ at $k^0=\gamma$. 

Thus the time integration selects $k^0=\gamma$ at $t \to \infty$. Consequently the exponential spatial suppression disappears, leaving
the long-range stationary contribution. This approximation itself 
is problematic near $u=0$ because $f(u+\gamma)$ is oscillating at $k=\sqrt{u}$, faster than $sin(ctu/2)/u$. Apart from this, physical constraints also raise incompatibility. The relevant issue is when contribution at distance $r=d$ becomes significant before factor 
\begin{equation}
    \frac{\sin\!\left[ct(k^0-\gamma)/2\right]}{k^0-\gamma}.
\end{equation}
peaks at $k^0=\gamma$.

At a specific point of integration, given by
\begin{equation}
    u=k^0-\gamma=\frac{p^2}{2mc\hbar},
\end{equation}
the phase in $f(k^0)$ and the factor leave the region of incoherent oscillation suppression and becomes coherent at order $p\cdot r\sim\pi\hbar$ as given by oscillatory phase in $f(k^0)\propto \cos(pr/\hbar)$; $\frac{p^2}{2m}\cdot t\sim\pi\hbar$ as given by oscillatory phase in $\sin(ctu/2)/u$. Eliminate $p$ to obtain:
\begin{equation}
    t\sim \frac{2mr^2}{\pi\hbar}
\end{equation}
This is the time when $r=d$ leaves the region of incoherent oscillation suppression and becomes coherent, before converging to the infinite time result as the factor peaks. Therefore, the amplitude at this order for $r=d$ is significant only:
\begin{equation}
    t\gtrsim \frac{md^2}{\hbar}
\end{equation}
In terms of wave-packet overlap measured by $\sigma_t$, Eq.~\eqref{eq:sigmat} gives
\begin{align}
t
\gtrsim
\frac{md^2}{\hbar}
\quad
&\Longrightarrow
\quad
\sigma_t^2
\ge
\frac{\hbar t}{m}
\gtrsim
d^2
\nonumber\\
&\Longrightarrow
\quad
\sigma_t
\gtrsim
d.
\label{eq:spread-contradiction}
\end{align}
However, two spatial modes can represent independently addressable subsystems only
while their overlap is negligible. For equal Gaussian modes $A,B$ of width $\sigma_t$
and center separation $d_{AB}$,
\begin{align}
\left|\braket{A(t)|B(t)}\right|
&=
\exp\!\left[
-\frac{d_{AB}^2}{4\sigma_t^2}
\right],
\end{align}
so the separated-mode regime for our system, where $A,B$ correspond to $1R,2L$, requires
\begin{equation}
\frac{d^2}{\sigma_t^2}
\gg
1.
\label{eq:distinct}
\end{equation}
Equations~\eqref{eq:spread-contradiction} and \eqref{eq:distinct} cannot hold
simultaneously. Hence the time needed to justify the stationary $1/d$ threshold
kernel is already a time at which the unbound packets do not define two separated
subsystems.

The same incompatibility follows directly from the finite-time propagation bound.
Using Eq.~\eqref{eq:rigorous-fixed-order}, significant amplitude imply
\begin{align}
\frac{md^2}{\hbar t}
&\sim
\frac{d^2}{\sigma_t^2}
\lesssim 
1.
\label{eq:eta}
\end{align}
Therefore every fixed perturbative order $n=O(1)$ in
Eq.~\eqref{eq:rigorous-fixed-order} is exponentially suppressed while the two
modes remain distinct. Increasing $n$ until the exponent becomes order unity
does not provide a perturbative loophole since the amplitude at order $n$ decrease exponentially in $n$. 

Finally, the discontinuous top-hat profile mentioned in the main text is not very relevant to the bound in this perturbative calculation, but it does have a pathodology. For
\begin{equation}
f(\bx)
=
\frac{\Theta(R-|\bx-\bX|)}{\sqrt V},
\qquad
V=\frac{4\pi R^3}{3},
\end{equation}
the formal gradient contains a delta distribution at the boundary,
\begin{equation}
\nabla f(\bx)
=
-\frac{\hat{\br}}{\sqrt V}\delta(r-R),
\end{equation}
and the free kinetic energy
\begin{equation}
\langle\hat T\rangle
=
\frac{\hbar^2}{2m}
\int\dd^3x\,|\nabla f|^2
\end{equation}
is ultraviolet divergent. Smoothing the boundary over a thickness $a$ gives
$|\nabla f|\sim(\sqrt V\,a)^{-1}$ in a shell of volume
$\sim4\pi R^2a$, and therefore
\begin{align}
\int\dd^3x\,|\nabla f|^2
&\sim
4\pi R^2a
\frac{1}{Va^2}
=
\frac{4\pi R^2}{Va}
\propto
\frac{1}{a}.
\end{align}
Thus the kinetic energy diverges as $a\to0$. For every physical $a>0$ the state
has a finite momentum spread and disperses under free evolution unless a
confining potential is supplied. With confinement, the appropriate matter line
is instead the evanescent propagator derived in
Eq.~\eqref{eq:Yukawa}.

Combining the four derivations, the exchange contribution in Eq.~(4) is always
bounded by an off-diagonal matter propagation amplitude. If the matter is
physically localized, each fixed-order off-diagonal contribution is exponentially attenuated as in Eq.~(5).
If it is genuinely unbound, the finite-time amplitude is propagation-suppressed
as in Eq.~(6) while the modes remain distinct. The apparent stationary
power-law channel is obtained only after the distributional limit in Eq.~(7) is
taken in a time regime incompatible with the same unbound packets remaining
spatially separated.

\clearpage
\twocolumngrid

\nocite{*}

\bibliography{apssamp}

\begin{thebibliography}{23}%
\makeatletter
\providecommand \@ifxundefined [1]{%
 \@ifx{#1\undefined}
}%
\providecommand \@ifnum [1]{%
 \ifnum #1\expandafter \@firstoftwo
 \else \expandafter \@secondoftwo
 \fi
}%
\providecommand \@ifx [1]{%
 \ifx #1\expandafter \@firstoftwo
 \else \expandafter \@secondoftwo
 \fi
}%
\providecommand \natexlab [1]{#1}%
\providecommand \enquote  [1]{``#1''}%
\providecommand \bibnamefont  [1]{#1}%
\providecommand \bibfnamefont [1]{#1}%
\providecommand \citenamefont [1]{#1}%
\providecommand \href@noop [0]{\@secondoftwo}%
\providecommand \href [0]{\begingroup \@sanitize@url \@href}%
\providecommand \@href[1]{\@@startlink{#1}\@@href}%
\providecommand \@@href[1]{\endgroup#1\@@endlink}%
\providecommand \@sanitize@url [0]{\catcode `\\12\catcode `\$12\catcode
  `\&12\catcode `\#12\catcode `\^12\catcode `\_12\catcode `\%12\relax}%
\providecommand \@@startlink[1]{}%
\providecommand \@@endlink[0]{}%
\providecommand \url  [0]{\begingroup\@sanitize@url \@url }%
\providecommand \@url [1]{\endgroup\@href {#1}{\urlprefix }}%
\providecommand \urlprefix  [0]{URL }%
\providecommand \Eprint [0]{\href }%
\providecommand \doibase [0]{https://doi.org/}%
\providecommand \selectlanguage [0]{\@gobble}%
\providecommand \bibinfo  [0]{\@secondoftwo}%
\providecommand \bibfield  [0]{\@secondoftwo}%
\providecommand \translation [1]{[#1]}%
\providecommand \BibitemOpen [0]{}%
\providecommand \bibitemStop [0]{}%
\providecommand \bibitemNoStop [0]{.\EOS\space}%
\providecommand \EOS [0]{\spacefactor3000\relax}%
\providecommand \BibitemShut  [1]{\csname bibitem#1\endcsname}%
\let\auto@bib@innerbib\@empty
\bibitem [{\citenamefont {Feynman}(2011)}]{Feynman1957}%
  \BibitemOpen
  \bibfield  {author} {\bibinfo {author} {\bibfnamefont {R.~P.}\ \bibnamefont
  {Feynman}},\ }\bibfield  {title} {\bibinfo {title} {The role of gravitation
  in physics: Report from the 1957 chapel hill conference},\ }in\ \href
  {https://pure.mpg.de/rest/items/item_2279905/component/file_3529186/content}
  {\emph {\bibinfo {booktitle} {Proceedings of the 1957 Chapel Hill
  Conference}}},\ \bibinfo {editor} {edited by\ \bibinfo {editor}
  {\bibfnamefont {C.~M.}\ \bibnamefont {DeWitt}}\ and\ \bibinfo {editor}
  {\bibfnamefont {D.}~\bibnamefont {Rickles}}}\ (\bibinfo  {publisher} {Max
  Planck Research Library for the History and Development of Knowledge},\
  \bibinfo {address} {Berlin},\ \bibinfo {year} {2011})\ \bibinfo {note}
  {originally published in 1957}\BibitemShut {NoStop}%
\bibitem [{\citenamefont {Bose}\ \emph {et~al.}(2017)\citenamefont {Bose},
  \citenamefont {Mazumdar}, \citenamefont {Morley}, \citenamefont {Ulbricht},
  \citenamefont {Toro\ifmmode~\check{s}\else \v{s}\fi{}}, \citenamefont
  {Paternostro}, \citenamefont {Geraci}, \citenamefont {Barker}, \citenamefont
  {Kim},\ and\ \citenamefont {Milburn}}]{PhysRevLett.119.240401}%
  \BibitemOpen
  \bibfield  {author} {\bibinfo {author} {\bibfnamefont {S.}~\bibnamefont
  {Bose}}, \bibinfo {author} {\bibfnamefont {A.}~\bibnamefont {Mazumdar}},
  \bibinfo {author} {\bibfnamefont {G.~W.}\ \bibnamefont {Morley}}, \bibinfo
  {author} {\bibfnamefont {H.}~\bibnamefont {Ulbricht}}, \bibinfo {author}
  {\bibfnamefont {M.}~\bibnamefont {Toro\ifmmode~\check{s}\else \v{s}\fi{}}},
  \bibinfo {author} {\bibfnamefont {M.}~\bibnamefont {Paternostro}}, \bibinfo
  {author} {\bibfnamefont {A.~A.}\ \bibnamefont {Geraci}}, \bibinfo {author}
  {\bibfnamefont {P.~F.}\ \bibnamefont {Barker}}, \bibinfo {author}
  {\bibfnamefont {M.~S.}\ \bibnamefont {Kim}},\ and\ \bibinfo {author}
  {\bibfnamefont {G.}~\bibnamefont {Milburn}},\ }\bibfield  {title} {\bibinfo
  {title} {Spin entanglement witness for quantum gravity},\ }\href
  {https://doi.org/10.1103/PhysRevLett.119.240401} {\bibfield  {journal}
  {\bibinfo  {journal} {Phys. Rev. Lett.}\ }\textbf {\bibinfo {volume} {119}},\
  \bibinfo {pages} {240401} (\bibinfo {year} {2017})}\BibitemShut {NoStop}%
\bibitem [{\citenamefont {Marletto}\ and\ \citenamefont
  {Vedral}(2017)}]{PhysRevLett.119.240402}%
  \BibitemOpen
  \bibfield  {author} {\bibinfo {author} {\bibfnamefont {C.}~\bibnamefont
  {Marletto}}\ and\ \bibinfo {author} {\bibfnamefont {V.}~\bibnamefont
  {Vedral}},\ }\bibfield  {title} {\bibinfo {title} {Gravitationally induced
  entanglement between two massive particles is sufficient evidence of quantum
  effects in gravity},\ }\href {https://doi.org/10.1103/PhysRevLett.119.240402}
  {\bibfield  {journal} {\bibinfo  {journal} {Phys. Rev. Lett.}\ }\textbf
  {\bibinfo {volume} {119}},\ \bibinfo {pages} {240402} (\bibinfo {year}
  {2017})}\BibitemShut {NoStop}%
\bibitem [{\citenamefont {Krisnanda}\ \emph {et~al.}(2020)\citenamefont
  {Krisnanda}, \citenamefont {Tham}, \citenamefont {Paternostro},\ and\
  \citenamefont {Paterek}}]{krisnanda2020observable}%
  \BibitemOpen
  \bibfield  {author} {\bibinfo {author} {\bibfnamefont {T.}~\bibnamefont
  {Krisnanda}}, \bibinfo {author} {\bibfnamefont {G.~Y.}\ \bibnamefont {Tham}},
  \bibinfo {author} {\bibfnamefont {M.}~\bibnamefont {Paternostro}},\ and\
  \bibinfo {author} {\bibfnamefont {T.}~\bibnamefont {Paterek}},\ }\bibfield
  {title} {\bibinfo {title} {Observable quantum entanglement due to gravity},\
  }\href {https://www.nature.com/articles/s41534-020-0243-y} {\bibfield
  {journal} {\bibinfo  {journal} {npj Quantum Inf.}\ }\textbf {\bibinfo
  {volume} {6}},\ \bibinfo {pages} {12} (\bibinfo {year} {2020})}\BibitemShut
  {NoStop}%
\bibitem [{\citenamefont {Biswas}\ \emph {et~al.}(2023)\citenamefont {Biswas},
  \citenamefont {Bose}, \citenamefont {Mazumdar},\ and\ \citenamefont
  {Toroš}}]{Biswas2023}%
  \BibitemOpen
  \bibfield  {author} {\bibinfo {author} {\bibfnamefont {D.}~\bibnamefont
  {Biswas}}, \bibinfo {author} {\bibfnamefont {S.}~\bibnamefont {Bose}},
  \bibinfo {author} {\bibfnamefont {A.}~\bibnamefont {Mazumdar}},\ and\
  \bibinfo {author} {\bibfnamefont {M.}~\bibnamefont {Toroš}},\ }\bibfield
  {title} {\bibinfo {title} {Gravitational optomechanics: Photon-matter
  entanglement via graviton exchange},\ }\bibfield  {journal} {\bibinfo
  {journal} {Physical Review D}\ }\textbf {\bibinfo {volume} {108}},\ \href
  {https://doi.org/10.1103/physrevd.108.064023} {10.1103/physrevd.108.064023}
  (\bibinfo {year} {2023})\BibitemShut {NoStop}%
\bibitem [{\citenamefont {Bose}\ \emph {et~al.}(2025)\citenamefont {Bose},
  \citenamefont {Fuentes}, \citenamefont {Geraci}, \citenamefont {Khan},
  \citenamefont {Qvarfort}, \citenamefont {Rademacher}, \citenamefont {Rashid},
  \citenamefont {Toro{\v{s}}}, \citenamefont {Ulbricht},\ and\ \citenamefont
  {Wanjura}}]{bose2025massive}%
  \BibitemOpen
  \bibfield  {author} {\bibinfo {author} {\bibfnamefont {S.}~\bibnamefont
  {Bose}}, \bibinfo {author} {\bibfnamefont {I.}~\bibnamefont {Fuentes}},
  \bibinfo {author} {\bibfnamefont {A.~A.}\ \bibnamefont {Geraci}}, \bibinfo
  {author} {\bibfnamefont {S.~M.}\ \bibnamefont {Khan}}, \bibinfo {author}
  {\bibfnamefont {S.}~\bibnamefont {Qvarfort}}, \bibinfo {author}
  {\bibfnamefont {M.}~\bibnamefont {Rademacher}}, \bibinfo {author}
  {\bibfnamefont {M.}~\bibnamefont {Rashid}}, \bibinfo {author} {\bibfnamefont
  {M.}~\bibnamefont {Toro{\v{s}}}}, \bibinfo {author} {\bibfnamefont
  {H.}~\bibnamefont {Ulbricht}},\ and\ \bibinfo {author} {\bibfnamefont
  {C.~C.}\ \bibnamefont {Wanjura}},\ }\bibfield  {title} {\bibinfo {title}
  {Massive quantum systems as interfaces of quantum mechanics and gravity},\
  }\href {https://journals.aps.org/rmp/abstract/10.1103/RevModPhys.97.015003}
  {\bibfield  {journal} {\bibinfo  {journal} {Rev. Mod. Phys.}\ }\textbf
  {\bibinfo {volume} {97}},\ \bibinfo {pages} {015003} (\bibinfo {year}
  {2025})}\BibitemShut {NoStop}%
\bibitem [{\citenamefont {Marletto}\ and\ \citenamefont
  {Vedral}(2025{\natexlab{a}})}]{marletto2025quantum}%
  \BibitemOpen
  \bibfield  {author} {\bibinfo {author} {\bibfnamefont {C.}~\bibnamefont
  {Marletto}}\ and\ \bibinfo {author} {\bibfnamefont {V.}~\bibnamefont
  {Vedral}},\ }\bibfield  {title} {\bibinfo {title} {Quantum-information
  methods for quantum gravity laboratory-based tests},\ }\href
  {https://journals.aps.org/rmp/abstract/10.1103/RevModPhys.97.015006}
  {\bibfield  {journal} {\bibinfo  {journal} {Rev. Mod. Phys.}\ }\textbf
  {\bibinfo {volume} {97}},\ \bibinfo {pages} {015006} (\bibinfo {year}
  {2025}{\natexlab{a}})}\BibitemShut {NoStop}%
\bibitem [{\citenamefont {Aziz}\ and\ \citenamefont
  {Howl}(2025)}]{aziz2025classical}%
  \BibitemOpen
  \bibfield  {author} {\bibinfo {author} {\bibfnamefont {J.}~\bibnamefont
  {Aziz}}\ and\ \bibinfo {author} {\bibfnamefont {R.}~\bibnamefont {Howl}},\
  }\bibfield  {title} {\bibinfo {title} {Classical theories of gravity produce
  entanglement},\ }\href {https://doi.org/10.1038/s41586-025-09595-7}
  {\bibfield  {journal} {\bibinfo  {journal} {Nature}\ }\textbf {\bibinfo
  {volume} {646}},\ \bibinfo {pages} {813} (\bibinfo {year}
  {2025})}\BibitemShut {NoStop}%
\bibitem [{\citenamefont {Diósi}(2026)}]{diosi2025no}%
  \BibitemOpen
  \bibfield  {author} {\bibinfo {author} {\bibfnamefont {L.}~\bibnamefont
  {Diósi}},\ }\href {https://arxiv.org/abs/2511.00852} {\bibinfo {title} {No,
  classical gravity does not entangle quantized matter fields}} (\bibinfo
  {year} {2026}),\ \Eprint {https://arxiv.org/abs/2511.00852} {arXiv:2511.00852
  [quant-ph]} \BibitemShut {NoStop}%
\bibitem [{\citenamefont {Marletto}\ \emph {et~al.}(2025)\citenamefont
  {Marletto}, \citenamefont {Oppenheim}, \citenamefont {Vedral},\ and\
  \citenamefont {Wilson}}]{marletto2025classical}%
  \BibitemOpen
  \bibfield  {author} {\bibinfo {author} {\bibfnamefont {C.}~\bibnamefont
  {Marletto}}, \bibinfo {author} {\bibfnamefont {J.}~\bibnamefont {Oppenheim}},
  \bibinfo {author} {\bibfnamefont {V.}~\bibnamefont {Vedral}},\ and\ \bibinfo
  {author} {\bibfnamefont {E.}~\bibnamefont {Wilson}},\ }\href
  {https://arxiv.org/abs/2511.07348} {\bibinfo {title} {Classical gravity
  cannot mediate entanglement}} (\bibinfo {year} {2025}),\ \Eprint
  {https://arxiv.org/abs/2511.07348} {arXiv:2511.07348 [quant-ph]} \BibitemShut
  {NoStop}%
\bibitem [{\citenamefont {Gundhi}\ \emph {et~al.}(2026)\citenamefont {Gundhi},
  \citenamefont {Infantino},\ and\ \citenamefont
  {Bassi}}]{gundhi2026classical}%
  \BibitemOpen
  \bibfield  {author} {\bibinfo {author} {\bibfnamefont {A.}~\bibnamefont
  {Gundhi}}, \bibinfo {author} {\bibfnamefont {G.}~\bibnamefont {Infantino}},\
  and\ \bibinfo {author} {\bibfnamefont {A.}~\bibnamefont {Bassi}},\ }\href
  {https://arxiv.org/abs/2604.19696} {\bibinfo {title} {Can classical theories
  of gravity produce entanglement?}} (\bibinfo {year} {2026}),\ \Eprint
  {https://arxiv.org/abs/2604.19696} {arXiv:2604.19696 [quant-ph]} \BibitemShut
  {NoStop}%
\bibitem [{\citenamefont {Marletto}\ and\ \citenamefont
  {Vedral}(2025{\natexlab{b}})}]{marletto2025local}%
  \BibitemOpen
  \bibfield  {author} {\bibinfo {author} {\bibfnamefont {C.}~\bibnamefont
  {Marletto}}\ and\ \bibinfo {author} {\bibfnamefont {V.}~\bibnamefont
  {Vedral}},\ }\href {https://arxiv.org/abs/2510.19969} {\bibinfo {title}
  {Classical gravity cannot mediate entanglement by local means}} (\bibinfo
  {year} {2025}{\natexlab{b}}),\ \Eprint {https://arxiv.org/abs/2510.19969}
  {arXiv:2510.19969 [quant-ph]} \BibitemShut {NoStop}%
\bibitem [{\citenamefont {Schneider}\ \emph {et~al.}(2026)\citenamefont
  {Schneider}, \citenamefont {Huggett},\ and\ \citenamefont
  {Linnemann}}]{schneider2025demonstration}%
  \BibitemOpen
  \bibfield  {author} {\bibinfo {author} {\bibfnamefont {M.~D.}\ \bibnamefont
  {Schneider}}, \bibinfo {author} {\bibfnamefont {N.}~\bibnamefont {Huggett}},\
  and\ \bibinfo {author} {\bibfnamefont {N.}~\bibnamefont {Linnemann}},\ }\href
  {https://arxiv.org/abs/2511.19242} {\bibinfo {title} {A demonstration that
  classical gravity does not produce entanglement}} (\bibinfo {year} {2026}),\
  \Eprint {https://arxiv.org/abs/2511.19242} {arXiv:2511.19242 [quant-ph]}
  \BibitemShut {NoStop}%
\bibitem [{\citenamefont {Vidal}\ and\ \citenamefont
  {Iyer}(2026)}]{vidal2026matter}%
  \BibitemOpen
  \bibfield  {author} {\bibinfo {author} {\bibfnamefont {N.~T.}\ \bibnamefont
  {Vidal}}\ and\ \bibinfo {author} {\bibfnamefont {A.~V.}\ \bibnamefont
  {Iyer}},\ }\href {https://arxiv.org/abs/2607.03429} {\bibinfo {title}
  {Matter-field exchange generates entanglement, not classical gravity}}
  (\bibinfo {year} {2026}),\ \Eprint {https://arxiv.org/abs/2607.03429}
  {arXiv:2607.03429 [quant-ph]} \BibitemShut {NoStop}%
\bibitem [{\citenamefont {van~de Kamp}\ \emph {et~al.}(2020)\citenamefont
  {van~de Kamp}, \citenamefont {Marshman}, \citenamefont {Bose},\ and\
  \citenamefont {Mazumdar}}]{PhysRevA.102.062807}%
  \BibitemOpen
  \bibfield  {author} {\bibinfo {author} {\bibfnamefont {T.~W.}\ \bibnamefont
  {van~de Kamp}}, \bibinfo {author} {\bibfnamefont {R.~J.}\ \bibnamefont
  {Marshman}}, \bibinfo {author} {\bibfnamefont {S.}~\bibnamefont {Bose}},\
  and\ \bibinfo {author} {\bibfnamefont {A.}~\bibnamefont {Mazumdar}},\
  }\bibfield  {title} {\bibinfo {title} {Quantum gravity witness via
  entanglement of masses: Casimir screening},\ }\href
  {https://doi.org/10.1103/PhysRevA.102.062807} {\bibfield  {journal} {\bibinfo
   {journal} {Phys. Rev. A}\ }\textbf {\bibinfo {volume} {102}},\ \bibinfo
  {pages} {062807} (\bibinfo {year} {2020})}\BibitemShut {NoStop}%
\bibitem [{\citenamefont {Toro\ifmmode~\check{s}\else \v{s}\fi{}}\ \emph
  {et~al.}(2021)\citenamefont {Toro\ifmmode~\check{s}\else \v{s}\fi{}},
  \citenamefont {van~de Kamp}, \citenamefont {Marshman}, \citenamefont {Kim},
  \citenamefont {Mazumdar},\ and\ \citenamefont
  {Bose}}]{PhysRevResearch.3.023178}%
  \BibitemOpen
  \bibfield  {author} {\bibinfo {author} {\bibfnamefont {M.}~\bibnamefont
  {Toro\ifmmode~\check{s}\else \v{s}\fi{}}}, \bibinfo {author} {\bibfnamefont
  {T.~W.}\ \bibnamefont {van~de Kamp}}, \bibinfo {author} {\bibfnamefont
  {R.~J.}\ \bibnamefont {Marshman}}, \bibinfo {author} {\bibfnamefont {M.~S.}\
  \bibnamefont {Kim}}, \bibinfo {author} {\bibfnamefont {A.}~\bibnamefont
  {Mazumdar}},\ and\ \bibinfo {author} {\bibfnamefont {S.}~\bibnamefont
  {Bose}},\ }\bibfield  {title} {\bibinfo {title} {Relative acceleration noise
  mitigation for nanocrystal matter-wave interferometry: Applications to
  entangling masses via quantum gravity},\ }\href
  {https://doi.org/10.1103/PhysRevResearch.3.023178} {\bibfield  {journal}
  {\bibinfo  {journal} {Phys. Rev. Res.}\ }\textbf {\bibinfo {volume} {3}},\
  \bibinfo {pages} {023178} (\bibinfo {year} {2021})}\BibitemShut {NoStop}%
\bibitem [{\citenamefont {Matsumura}\ and\ \citenamefont
  {Yamamoto}(2020)}]{PhysRevD.102.106021}%
  \BibitemOpen
  \bibfield  {author} {\bibinfo {author} {\bibfnamefont {A.}~\bibnamefont
  {Matsumura}}\ and\ \bibinfo {author} {\bibfnamefont {K.}~\bibnamefont
  {Yamamoto}},\ }\bibfield  {title} {\bibinfo {title} {Gravity-induced
  entanglement in optomechanical systems},\ }\href
  {https://doi.org/10.1103/PhysRevD.102.106021} {\bibfield  {journal} {\bibinfo
   {journal} {Phys. Rev. D}\ }\textbf {\bibinfo {volume} {102}},\ \bibinfo
  {pages} {106021} (\bibinfo {year} {2020})}\BibitemShut {NoStop}%
\bibitem [{\citenamefont {Miki}\ \emph {et~al.}(2024)\citenamefont {Miki},
  \citenamefont {Matsumura},\ and\ \citenamefont
  {Yamamoto}}]{PhysRevD.110.024057}%
  \BibitemOpen
  \bibfield  {author} {\bibinfo {author} {\bibfnamefont {D.}~\bibnamefont
  {Miki}}, \bibinfo {author} {\bibfnamefont {A.}~\bibnamefont {Matsumura}},\
  and\ \bibinfo {author} {\bibfnamefont {K.}~\bibnamefont {Yamamoto}},\
  }\bibfield  {title} {\bibinfo {title} {Feasible generation of gravity-induced
  entanglement by using optomechanical systems},\ }\href
  {https://doi.org/10.1103/PhysRevD.110.024057} {\bibfield  {journal} {\bibinfo
   {journal} {Phys. Rev. D}\ }\textbf {\bibinfo {volume} {110}},\ \bibinfo
  {pages} {024057} (\bibinfo {year} {2024})}\BibitemShut {NoStop}%
\bibitem [{\citenamefont {Yant}\ and\ \citenamefont
  {Blencowe}(2023)}]{PhysRevD.107.106018}%
  \BibitemOpen
  \bibfield  {author} {\bibinfo {author} {\bibfnamefont {J.}~\bibnamefont
  {Yant}}\ and\ \bibinfo {author} {\bibfnamefont {M.}~\bibnamefont
  {Blencowe}},\ }\bibfield  {title} {\bibinfo {title} {Gravitationally induced
  entanglement in a harmonic trap},\ }\href
  {https://doi.org/10.1103/PhysRevD.107.106018} {\bibfield  {journal} {\bibinfo
   {journal} {Phys. Rev. D}\ }\textbf {\bibinfo {volume} {107}},\ \bibinfo
  {pages} {106018} (\bibinfo {year} {2023})}\BibitemShut {NoStop}%
\bibitem [{\citenamefont {Agafonova}\ \emph {et~al.}(2026)\citenamefont
  {Agafonova}, \citenamefont {Rossell{\'o}}, \citenamefont {Mekonnen},\ and\
  \citenamefont {Hosten}}]{agafonova2026one}%
  \BibitemOpen
  \bibfield  {author} {\bibinfo {author} {\bibfnamefont {S.}~\bibnamefont
  {Agafonova}}, \bibinfo {author} {\bibfnamefont {P.}~\bibnamefont
  {Rossell{\'o}}}, \bibinfo {author} {\bibfnamefont {M.}~\bibnamefont
  {Mekonnen}},\ and\ \bibinfo {author} {\bibfnamefont {O.}~\bibnamefont
  {Hosten}},\ }\bibfield  {title} {\bibinfo {title} {One-milligram torsional
  pendulum toward experiments at the quantum-gravity interface},\ }\href@noop
  {} {\bibfield  {journal} {\bibinfo  {journal} {Communications Physics}\
  }\textbf {\bibinfo {volume} {9}},\ \bibinfo {pages} {80} (\bibinfo {year}
  {2026})}\BibitemShut {NoStop}%
\bibitem [{\citenamefont {Carney}\ \emph {et~al.}(2021)\citenamefont {Carney},
  \citenamefont {M{\"u}ller},\ and\ \citenamefont {Taylor}}]{carney2021using}%
  \BibitemOpen
  \bibfield  {author} {\bibinfo {author} {\bibfnamefont {D.}~\bibnamefont
  {Carney}}, \bibinfo {author} {\bibfnamefont {H.}~\bibnamefont {M{\"u}ller}},\
  and\ \bibinfo {author} {\bibfnamefont {J.~M.}\ \bibnamefont {Taylor}},\
  }\bibfield  {title} {\bibinfo {title} {Using an atom interferometer to infer
  gravitational entanglement generation},\ }\href@noop {} {\bibfield  {journal}
  {\bibinfo  {journal} {PRX Quantum}\ }\textbf {\bibinfo {volume} {2}},\
  \bibinfo {pages} {030330} (\bibinfo {year} {2021})}\BibitemShut {NoStop}%
\bibitem [{\citenamefont {Haine}(2021)}]{haine2021searching}%
  \BibitemOpen
  \bibfield  {author} {\bibinfo {author} {\bibfnamefont {S.~A.}\ \bibnamefont
  {Haine}},\ }\bibfield  {title} {\bibinfo {title} {Searching for signatures of
  quantum gravity in quantum gases},\ }\href@noop {} {\bibfield  {journal}
  {\bibinfo  {journal} {New Journal of Physics}\ }\textbf {\bibinfo {volume}
  {23}},\ \bibinfo {pages} {033020} (\bibinfo {year} {2021})}\BibitemShut
  {NoStop}%
\bibitem [{\citenamefont {Howl}\ \emph {et~al.}(2026)\citenamefont {Howl},
  \citenamefont {Cooper},\ and\ \citenamefont
  {Hackerm{\"u}ller}}]{howl2026gravitationally}%
  \BibitemOpen
  \bibfield  {author} {\bibinfo {author} {\bibfnamefont {R.}~\bibnamefont
  {Howl}}, \bibinfo {author} {\bibfnamefont {N.}~\bibnamefont {Cooper}},\ and\
  \bibinfo {author} {\bibfnamefont {L.}~\bibnamefont {Hackerm{\"u}ller}},\
  }\bibfield  {title} {\bibinfo {title} {Gravitationally induced entanglement
  in atom interferometry},\ }\href@noop {} {\bibfield  {journal} {\bibinfo
  {journal} {Physical Review A}\ }\textbf {\bibinfo {volume} {114}},\ \bibinfo
  {pages} {023306} (\bibinfo {year} {2026})}\BibitemShut {NoStop}%
\end{thebibliography}%

\end{document}